\documentclass[aps,twocolumn]{revtex4}
\usepackage{eurosym}
\usepackage{amsfonts}
\usepackage{amsmath}
\usepackage{amssymb,epsf}
\usepackage{color}
\usepackage{graphicx}
\usepackage{epstopdf}
\usepackage{float}
\usepackage{caption}
\usepackage{subfig}

\begin{document}

\title{Topological phantom AdS black holes in $F(R)$ gravity}
\author{B. Eslam Panah$^{1,2,3}$\footnote{
email address: eslampanah@umz.ac.ir}, and M. E. Rodrigues$^{4,5}$ \footnote{
email address: esialg@gmail.com}}
\affiliation{$^{1}$ Department of Theoretical Physics, Faculty of Science, University of
Mazandaran, P. O. Box 47416-95447, Babolsar, Iran}
\affiliation{$^{2}$ ICRANet-Mazandaran, University of Mazandaran, P. O. Box 47416-95447
Babolsar, Iran}
\affiliation{$^{3}$ ICRANet, Piazza della Repubblica 10, I-65122 Pescara, Italy}
\affiliation{$^{4}$ Faculdade de Ci\^{e}ncias Exatas e Tecnologia, Universidade Federal
do Par\'{a}\\
Campus Universit\'{a}rio de Abaetetuba, 68440-000, Abaetetuba, Par\'{a},
Brazil}
\affiliation{$^{5}$ Faculdade de F\'{\i}sica, Programa de P\'{o}s-Gradua\c{c}\~ao em
F\'isica, Universidade Federal do Par\'{a}, 66075-110, Bel\'{e}m, Par\'{a},
Brazil}

\begin{abstract}
In this paper, we obtain exact phantom (A)dS black hole solutions in the
context of $F(R)$ gravity with topological spacetime in four dimensions.
Then, we study the effects of different parameters on the event horizon. In
the following, we calculate the conserved and thermodynamic quantities of
the system and check the first law of thermodynamics for these kinds of
black holes. Next, we evaluate the local stability of the topological
phantom (A)dS black holes in $F(R)$ gravity by studying the heat capacity
and the geometrothemodynamic, where we show that the two approaches agrees.
We extend our study and investigate global stability by employing the Gibbs
potential and the Helmholtz free energy. In addition, the effects of
different parameters on local and global stabilities will be highlighted.
\end{abstract}

\maketitle





\section{introduction}


The observational evidence such as the luminosity distance of Supernovae
type Ia \cite{Perlmutter,PerlmutterI}, wide surveys on galaxies \cite{Gal},
and the anisotropy of cosmic microwave background radiation \cite{CMBR},
indicates that our Universe is currently undergoing a period of
acceleration. Identifying the cause of this late-time acceleration is a
challenging problem in cosmology. To describe this acceleration, some
candidates are proposed. One of the simplest ways to address this cosmic
acceleration is related to modifying the left-hand side of general
relativity (GR) field equations. This approach is known as the modified
theory of gravity. Among these modified theories of gravity, $F(R)$ gravity
includes some exciting features from both cosmological and astrophysical
points of view. The gravitational action in this modified theory of theory
is a general function of the scalar curvature $R$ \cite%
{F(R)I,F(R)II,F(R)III,F(R)IV}. This theory can be fixed according to the
astrophysical and cosmological observations \cite%
{Mod1,Mod2,Mod3,Mod4,Mod5,Mod6,Mod7}. Also, $F(R)$ gravity coincides with
Newtonian and post-Newtonian approximations \cite{CapozzielloI,CapozzielloII}%
. It may explain the structure formation of the Universe without considering
the dark matter. Moreover, the whole sequence of the Universe's evolution
epochs: inflation, radiation/matter dominance, and dark energy may be
extracted in $F(R)$ gravity. Another possibility to explain this accelerated
phase of our universe is the introduction of an exotic fluid called dark
energy \cite{dark}, where the fluid is isotropic with a negative pressure.
An alternative to describe this fluid is the phantom scalar field \cite%
{scalarphantom}, where the energy density is negative, so the pressure is
also negative, and can model dark energy.

On the other hand, black holes are exciting objects to study from
theoretical and observational points of view. To study the exciting
properties of black holes in any theory of gravity, we have to extract them.
According to the mentioned features of $F(R)$ gravity, we are interested to
extract black hole solutions in this theory. However, the field equations of 
$F(R)$ gravity are complicated fourth-order differential equations, and it
is not easy to find exact black hole solutions, especially in the present
matter field. Indeed, adding a matter field to $F(R)$\ gravity makes the
field equations much more difficult. However, some different (un)charged
black hole solutions in $F(R)$ gravity are obtained in Refs. \cite%
{BH1,BH2,BH3,BH4,BH5,BH6,BH7,BH8,BH9,BH10,BH11,BH12}.

The introduction of phantom fields to describe physical systems is old. In
1935, Einstein and Rose \cite{ER} introduced the so-called quasicharged
bridge, where the Reissner-Nordstr\"{o}m solution was used, but with a pure
imaginary charge, i.e. $q^{2}\rightarrow -q^{2}$, thus introducing a spin$-1$
phantom field as matter describing this gravitational configuration, even
though they did not show the action of the system. Then several situations
arose where the kinetic energy was negative or the field was phantom \cite%
{Visser}. Phantom black hole solutions are known in the literature \cite%
{PBH,PBH2,PBH3,PBH4}. Here, we want to investigate what the contribution to
the structure of the solution and its fundamental physical properties is
when we couple a spin$-1$ phantom field in a linear manner to the action of
the $F(R)$ theory in a topological metric.

The paper is divided as follows: first, in section \ref{sec1}, we establish
the equations of motion of the $F(R)$ theory, specify the case of constant
curvature, and obtain the solution. Subsequently, we define the essential
thermodynamic quantities in section \ref{sec2}. In section \ref{sec3}, we
study local and global thermodynamic stability as well as
geomtrothemodynamics. We make our concluding remarks in section \ref{sec4}.


\section{The field equations in \textbf{F(R) gravity and black hole solutions%
}}

\label{sec1} 

Here, we consider $F(R)$ gravity in which coupled with Maxwell field as a
matter source. The action of this theory in four-dimensional spacetime is
given by 
\begin{equation}
\mathcal{I}_{F(R)}=\int_{\partial \mathcal{M}}d^{4}x\sqrt{-g}\left[
F(R)+2\kappa ^{2}\eta \mathcal{F}\right] ,  \label{actionF(R)}
\end{equation}%
where the first term is related to the theory of $F(R)$ gravity in the form $%
F(R)=R+f\left( R\right) $, which $R$ is scalar curvature, and also, $f\left(
R\right) $ is an arbitrary function of scalar curvature $R$. In addition,
the second is the coupling with the Maxwell field, when $\eta =1$, or a
phantom field of spin $1$, when $\eta =-1$. It is notable that $\mathcal{F}=$%
\ $F_{\mu \nu }F^{\mu \nu }$ is the Maxwell invariant. Also, $F_{\mu \nu
}=\partial _{\mu }A_{\nu }-\partial _{\nu }A_{\mu }$ is the electromagnetic
tensor field, and $A_{\mu }$ is the gauge potential. Moreover, $\kappa
^{2}=8\pi G$, and $G$ is the Newtonian gravitational constant. In the above
action, $g=det(g_{\mu \nu })$ is the determinant of metric tensor $g_{\mu
\nu }$. Hereafter, we consider $G=c=1$.

We can obtain the equations of motion of $F(R)$ theory by varying the action
(\ref{actionF(R)}) with respect to the gravitational field $g_{\mu \nu }$,
and the gauge field $A_{\mu }$, lead to the following forms 
\begin{eqnarray}
R_{\mu \nu }\left( 1+f_{R}\right) -\frac{g_{\mu \nu }F(R)}{2}+\left( g_{\mu
\nu }\nabla ^{2}-\nabla _{\mu }\nabla _{\nu }\right) f_{R} &=&8\pi \mathrm{T}%
_{\mu \nu },  \label{EqF(R)1} \\
&&  \notag \\
\partial _{\mu }\left( \sqrt{-g}F^{\mu \nu }\right) &=&0,  \label{EqF(R)2}
\end{eqnarray}%
where $f_{R}=\frac{df(R)}{dR}$. Also, $\mathrm{T}_{\mu \nu }$ is the
energy-momentum tensor, and for four-dimensional spacetime can be written 
\begin{equation}
\mathrm{T}_{\mu \nu }=2\eta \left( \frac{1}{4}g_{\mu \nu }\mathcal{F}-F_{\mu
}^{~~\alpha }F_{\nu \alpha }\right) .  \label{Tmunu}
\end{equation}

We consider a topological four-dimensional static spacetime with the
following form 
\begin{equation}
ds^{2}=g(r)dt^{2}-\frac{dr^{2}}{g(r)}-r^{2}d\Omega _{k}^{2},  \label{Metric}
\end{equation}%
where $g(r)$ is the metric function. Also, in the above equation, $d\Omega
_{k}^{2}$ is given by 
\begin{equation}
d\Omega _{k}^{2}=\left\{ 
\begin{array}{ccc}
d\theta ^{2}+\sin ^{2}\theta d\varphi ^{2} &  & k=1 \\ 
d\theta ^{2}+d\varphi ^{2} &  & k=0 \\ 
d\theta ^{2}+\sinh ^{2}\theta d\varphi ^{2} &  & k=-1%
\end{array}%
\right. ,
\end{equation}

It is notable that the constant $k$ indicates that the boundary of $t=$
constant and $r=$constant can be elliptic ($k=1$), flat ($k=0$) or
hyperbolic ($k=-1$) curvature hypersurface.

In the following, we want to obtain the solutions for the constant scalar
curvature is $R=R_{0}=$ constant in four-dimensional spacetime. The trace of
Eq. (\ref{EqF(R)1}) yields 
\begin{equation}
R_{0}\left( 1+f_{R_{0}}\right) -2\left( R_{0}+f(R_{0})\right) =0,
\end{equation}%
where $f_{R_{0}}=$ $f_{R_{\left\vert _{R=R_{0}}\right. }}$, and the solution
for $R_{0}$ gives 
\begin{equation}
R_{0}=\frac{2f(R_{0})}{f_{R_{0}}-1}.  \label{R0}
\end{equation}

Substituting the equation (\ref{R0}) into Eq. (\ref{EqF(R)1}), we obtain the
equations of motion in $F(R)$-Maxwell (or phantom) theory which can be
written as 
\begin{equation}
R_{\mu \nu }\left( 1+f_{R_{0}}\right) -\frac{g_{\mu \nu }}{4}R_{0}\left(
1+f_{R_{0}}\right) =8\pi \mathrm{T}_{\mu \nu }.  \label{F(R)Trace}
\end{equation}

In order to obtain electrically charged black hole solutions, we consider a
radial electric field which its related gauge potential is in the following
form 
\begin{equation}
A_{\mu }=h\left( r\right) \delta _{\mu }^{t}
\end{equation}

We can find the following differential equation by using Eqs. (\ref{EqF(R)2}%
) and (\ref{Metric}) 
\begin{equation}
rh^{\prime \prime }(r)+2h^{\prime }(r)=0,  \label{hh}
\end{equation}%
where the prime and double prime are the first and the second derivatives
with respect to $r$, respectively. So the solution of the equation (\ref{hh}%
) is 
\begin{equation}
h(r)=\frac{q}{r},  \label{h(r)}
\end{equation}%
where $q$ is an integration constant which is related to the electric
charge. Considering the obtained $h(r)$ in Eq. (\ref{h(r)}), the
electromagnetic field tensor is given by 
\begin{equation}
F_{tr}=\partial _{t}A_{r}-\partial _{r}A_{t}=\frac{q}{r^{2}}.  \label{E(r)}
\end{equation}

Using the introduced metric (\ref{Metric}) and the field equations (\ref%
{F(R)Trace}), we want to obtain exact solutions for the metric function $%
g\left( r\right) $. After some calculation, we find the following
differential equations 
\begin{eqnarray}
eq_{tt} &=&eq_{rr}=2r^{3}\left( 1+f_{R_{0}}\right) \left( \frac{rR_{0}}{2}%
-rg^{\prime \prime }(r)-2g^{\prime }(r)\right)  \notag \\
&&-4\eta q^{2},  \label{eq11} \\
&&  \notag \\
eq_{\theta \theta } &=&eq_{\varphi \varphi }=4r^{2}\left( 1+f_{R_{0}}\right)
\left( g\left( r\right) -k-\frac{r^{2}R_{0}}{4}+rg^{\prime }(r)\right) 
\notag \\
&&+4\eta q^{2},  \label{eq22}
\end{eqnarray}%
where $eq_{tt}$, $eq_{rr}$, $eq_{\theta \theta }$ and $eq_{\varphi \varphi }$%
, respectively, are components of $tt$, $rr$, $\theta \theta $ and $\varphi
\varphi $ of field equations (\ref{F(R)Trace}). We are in a position to
obtain exact solutions for the constant scalar curvature ($R=R_{0}$= const).
We can extract the following metric function\ by considering Eqs. (\ref{eq11}%
) and (\ref{eq22}) as 
\begin{equation}
g(r)=k-\frac{2m}{r}+\frac{R_{0}r^{2}}{12}+\eta \frac{q^{2}}{\left(
1+f_{R_{0}}\right) r^{2}},  \label{g(r)F(R)}
\end{equation}%
where $m$ is an integration constant related to the total mass of the black
hole. The solution (\ref{g(r)F(R)}) satisfies all components of the field
equations (\ref{F(R)Trace}).

One of quantities that can give us information about the existence of
singularity is related to the Kretschmann scalar. Considering the
four-dimensional spacetime in Eq. (\ref{Metric}), with the metric function (%
\ref{g(r)F(R)}), we can obtain the Kretschmann scalar in the following form 
\begin{eqnarray}
R_{\alpha \beta \gamma \delta }R^{\alpha \beta \gamma \delta } &=&\frac{%
R_{0}^{2}}{6}+\frac{48m^{2}}{r^{6}}-\frac{96m\eta q^{2}}{\left(
1+f_{R_{0}}\right) r^{7}}  \notag \\
&&+\frac{56\eta ^{2}q^{4}}{\left( 1+f_{R_{0}}\right) ^{2}r^{8}},
\end{eqnarray}%
which indicates that the Kretschmann scalar diverges at $r=0$. In other
words, the Kretschmann scalar at $r\rightarrow 0$, leads to 
\begin{equation}
\lim_{r\longrightarrow 0}R_{\alpha \beta \gamma \delta }R^{\alpha \beta
\gamma \delta }\longrightarrow \infty .
\end{equation}%
So, there is a curvature singularity located at $r=0$. Also, it is finite
for $r\neq 0$.

The asymptotical behavior of the Kretschmann scalar is given by 
\begin{equation}
\lim_{r\longrightarrow \infty }R_{\alpha \beta \gamma \delta }R^{\alpha
\beta \gamma \delta }\longrightarrow \frac{R_{0}^{2}}{6},
\end{equation}%
also, the asymptotical behavior of the metric function leads to $%
\lim_{r\longrightarrow \infty }g\left( r\right) \longrightarrow \frac{%
R_{0}r^{2}}{12}$, which shows the spacetime will be asymptotically AdS, when
we define $R_{0}=-4\Lambda $. It is mentioned that we should restrict
ourselves to $f_{R}\neq -1$, to have physical solutions.

\begin{figure}[tbph]
\centering
\includegraphics[width=0.4\linewidth]{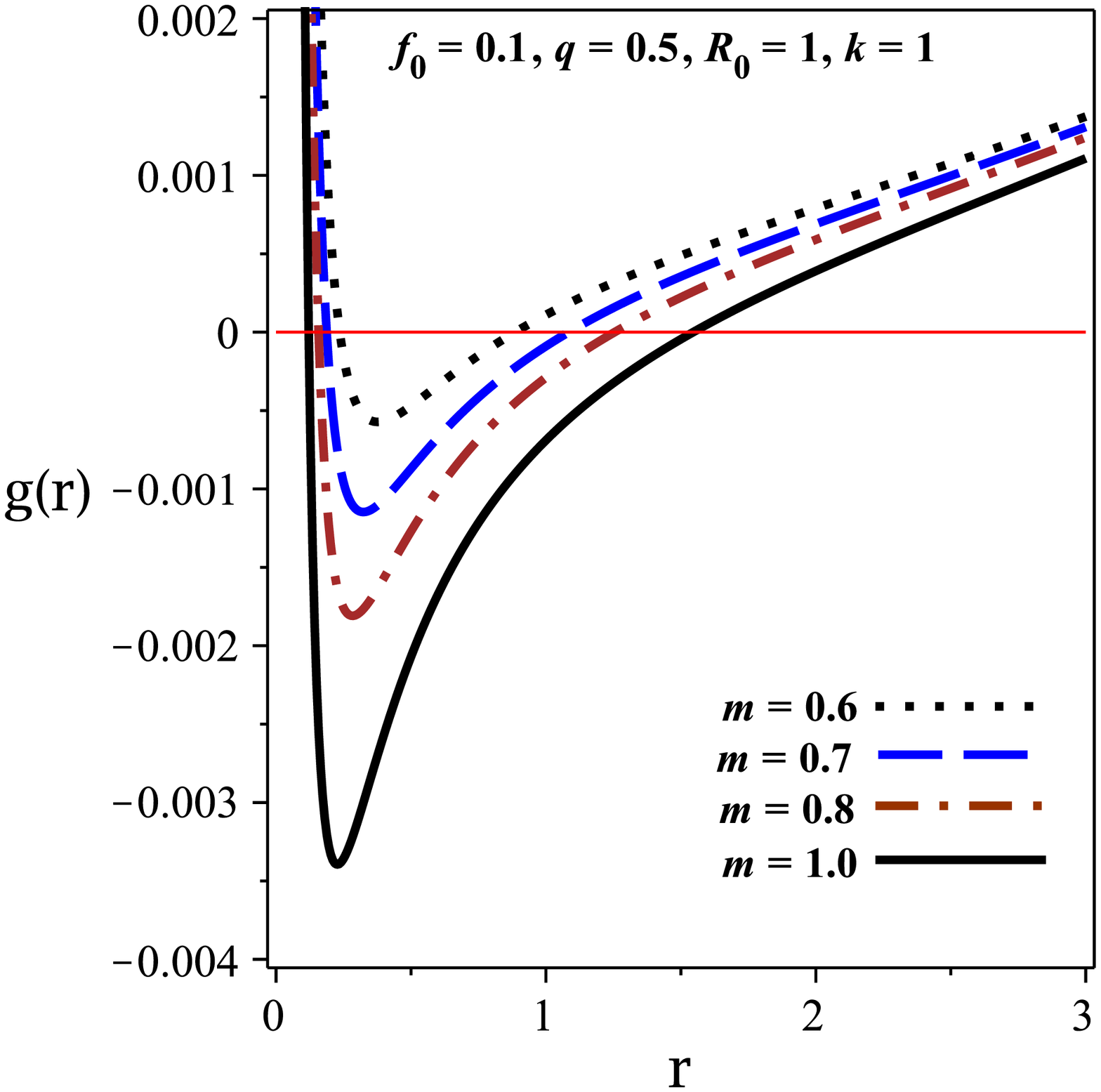} \includegraphics[width=0.4%
\linewidth]{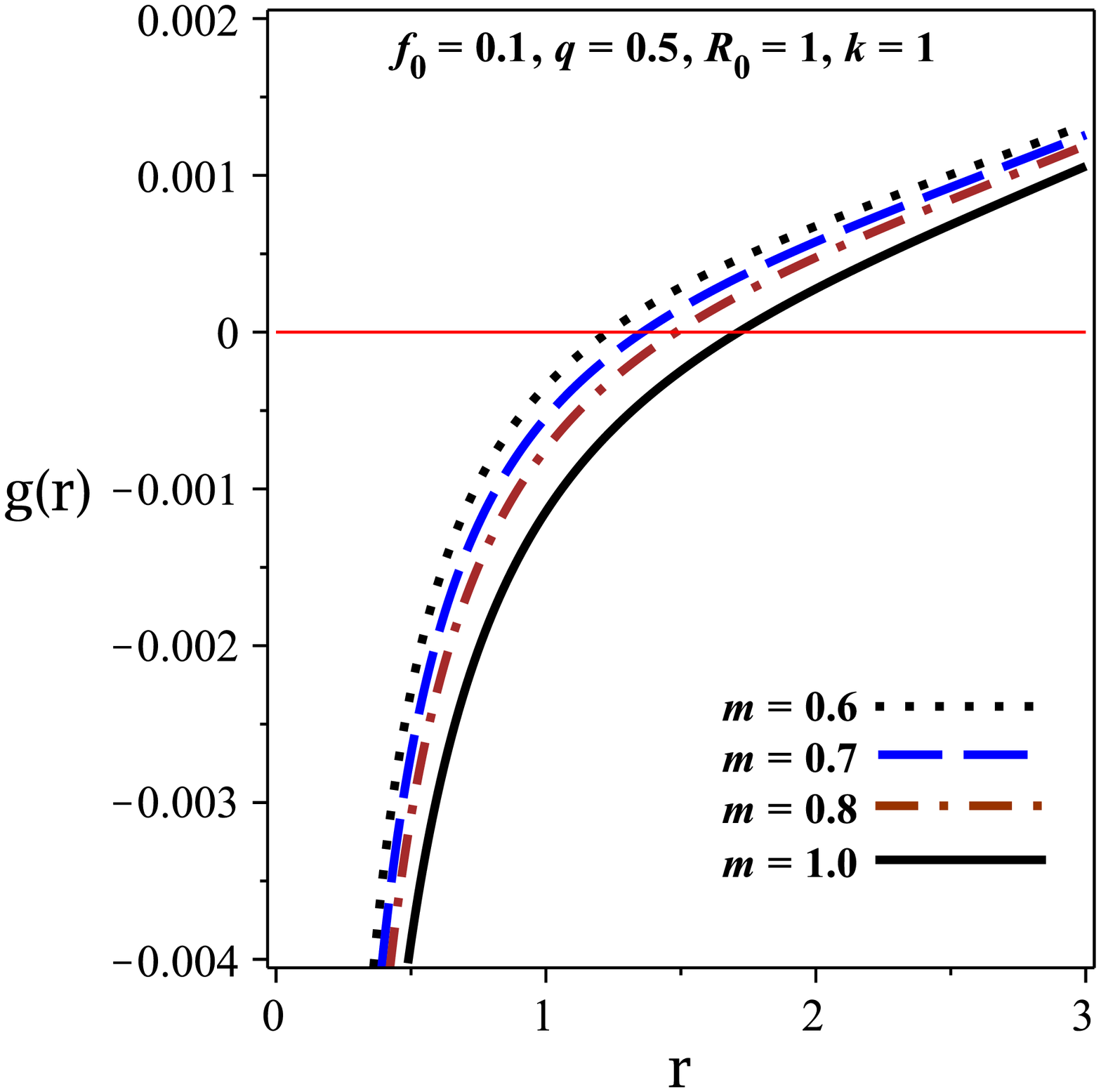}\newline
\includegraphics[width=0.4\linewidth]{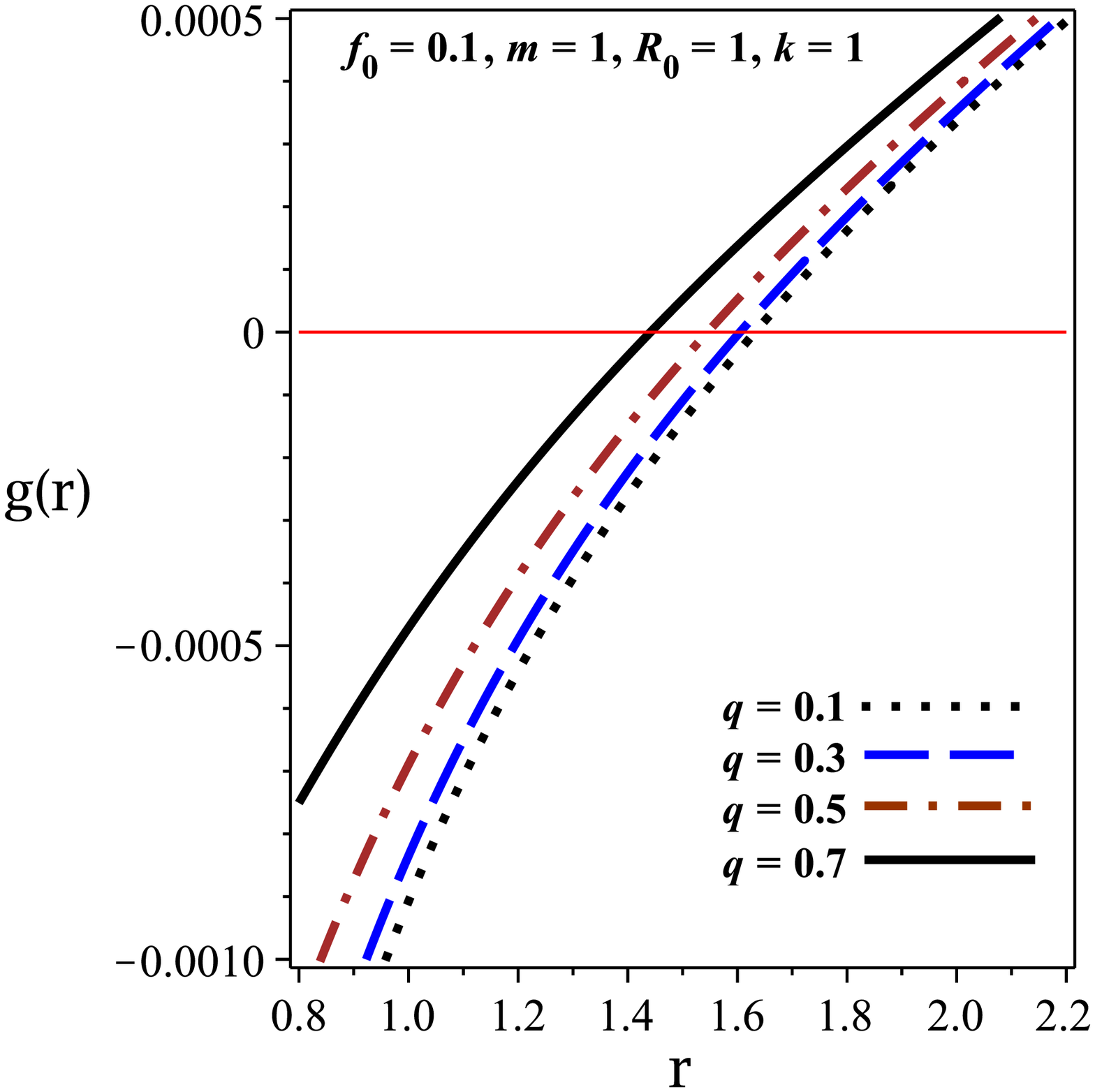} \includegraphics[width=0.4%
\linewidth]{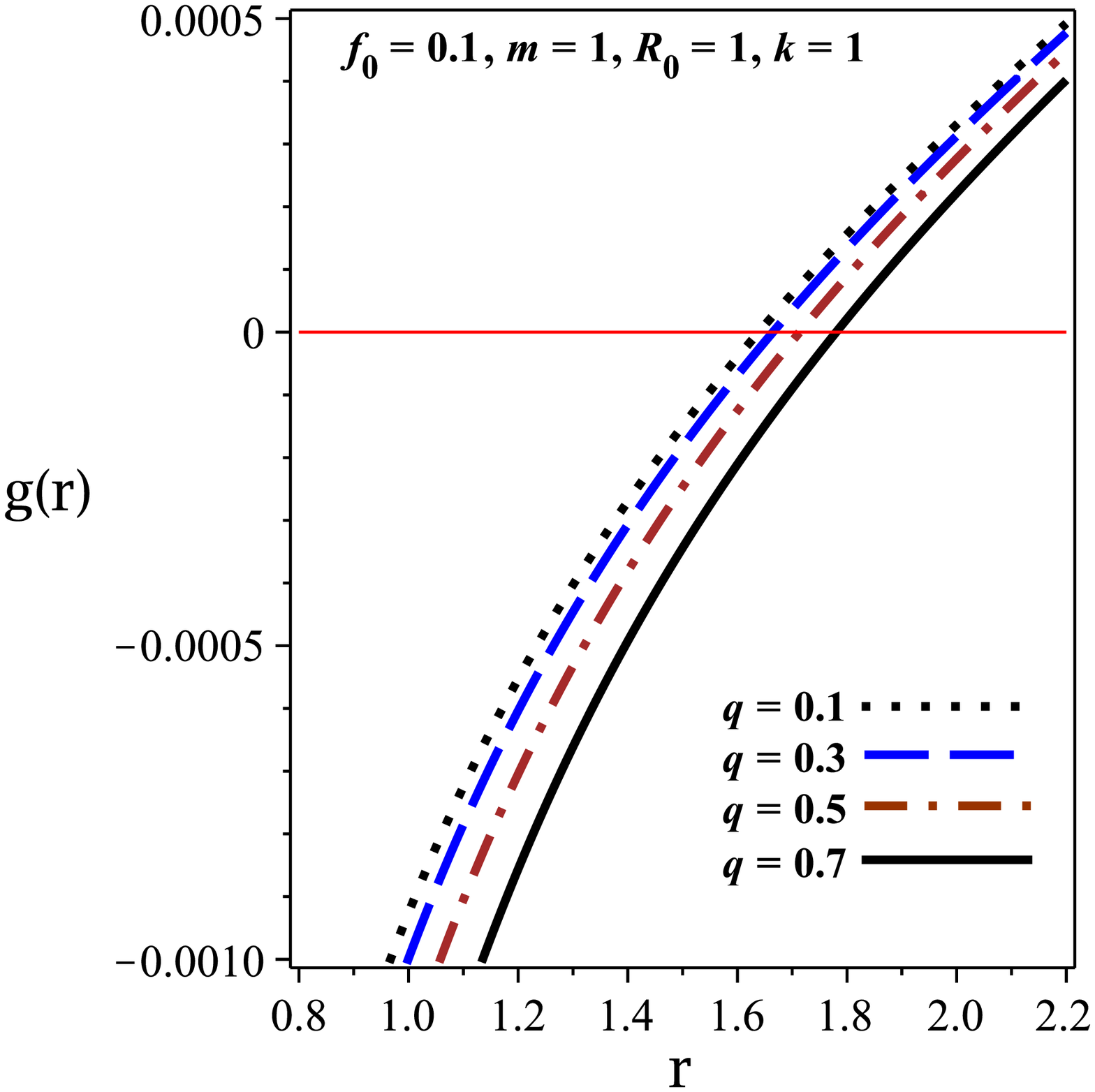}\newline
\includegraphics[width=0.4\linewidth]{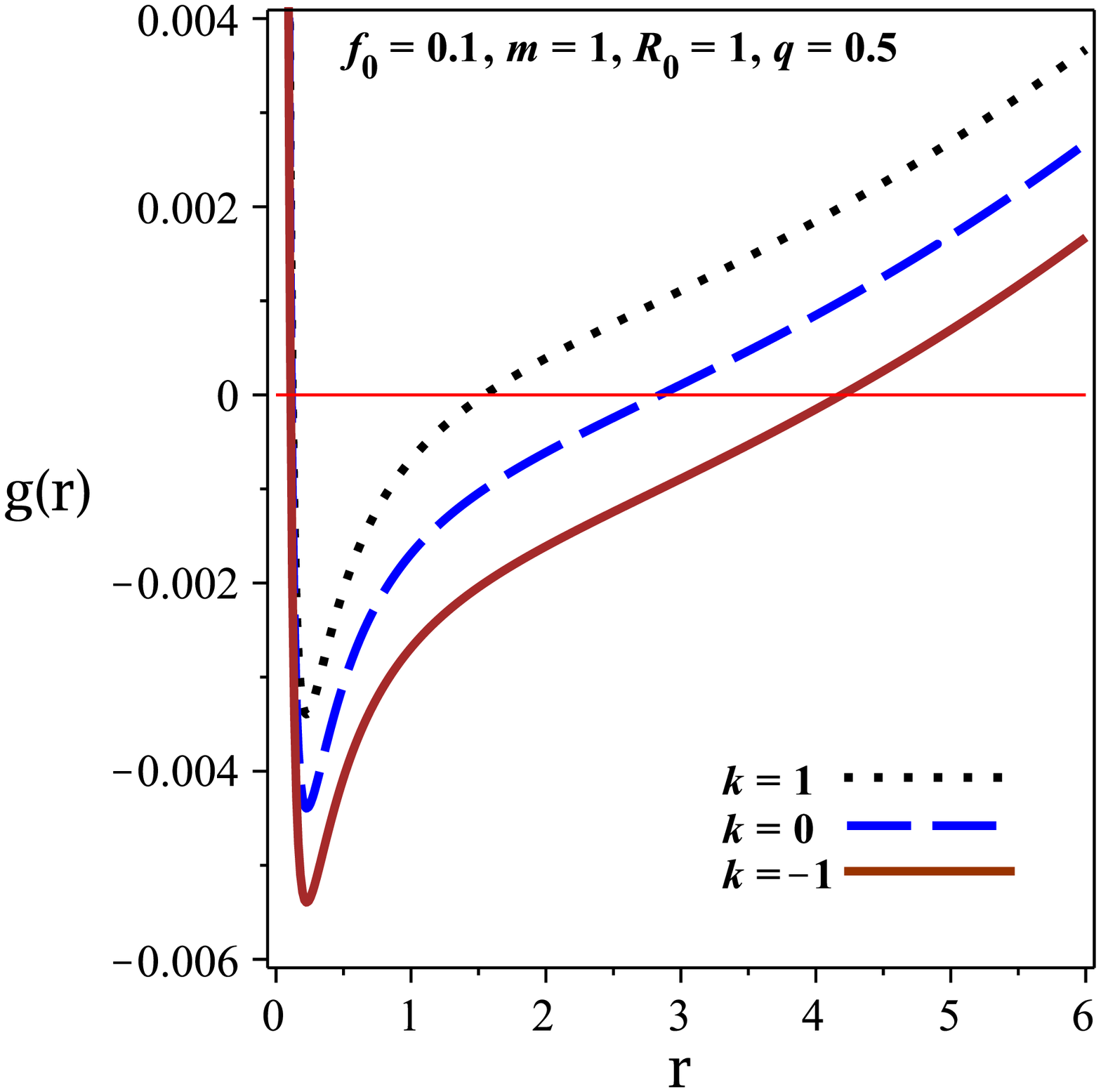} \includegraphics[width=0.4%
\linewidth]{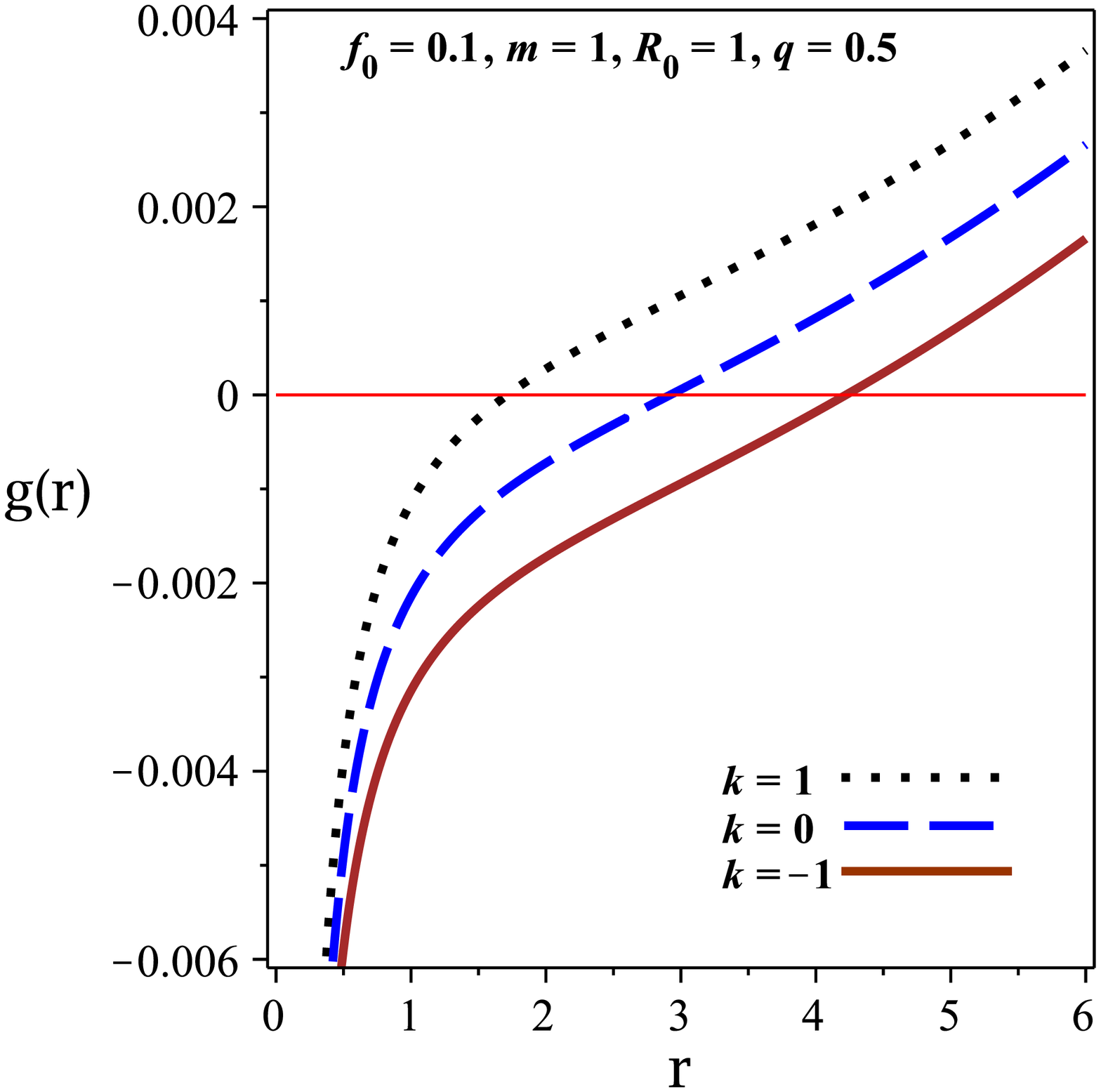}\newline
\includegraphics[width=0.4\linewidth]{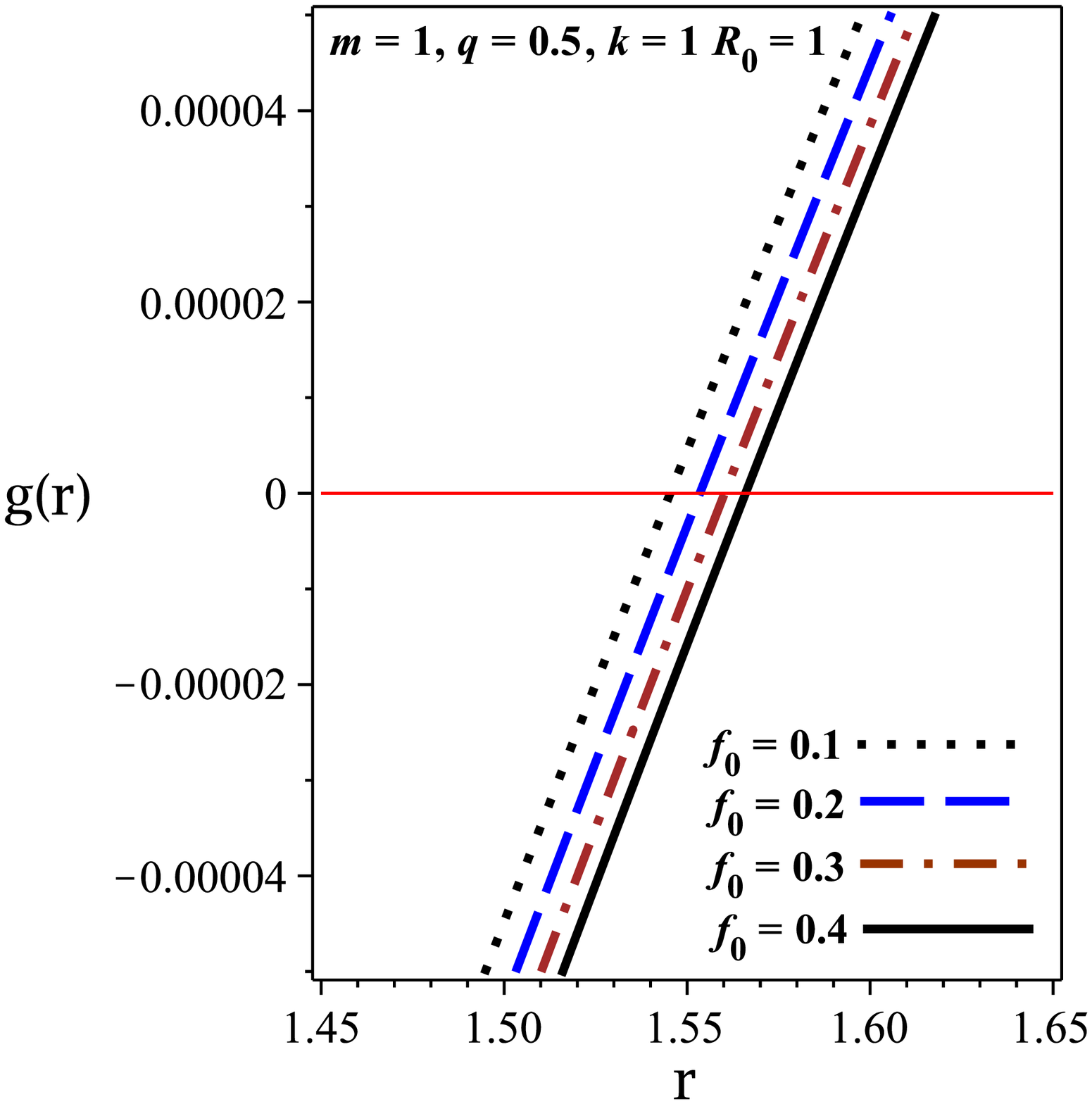} \includegraphics[width=0.4%
\linewidth]{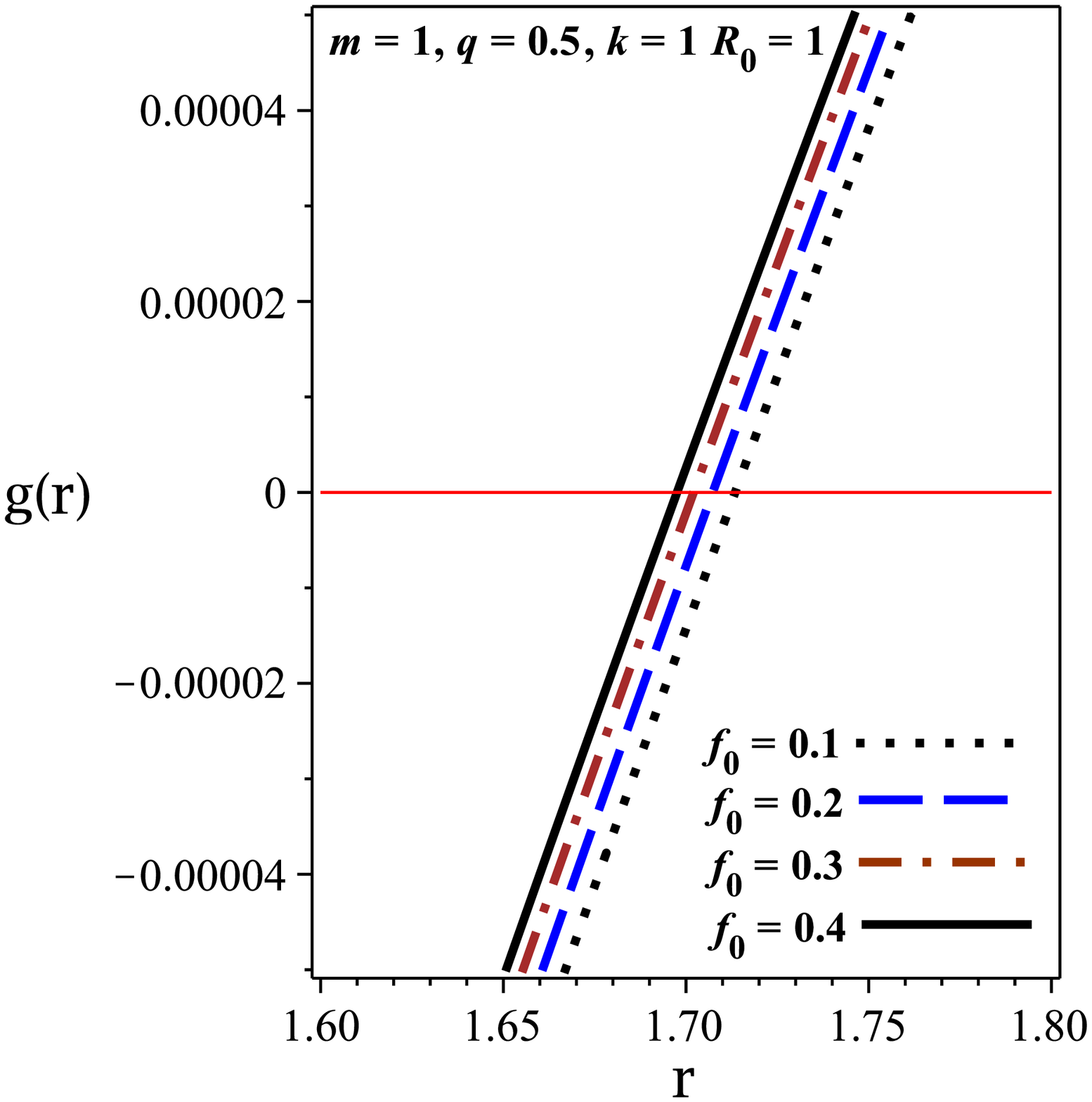}\newline
\includegraphics[width=0.4\linewidth]{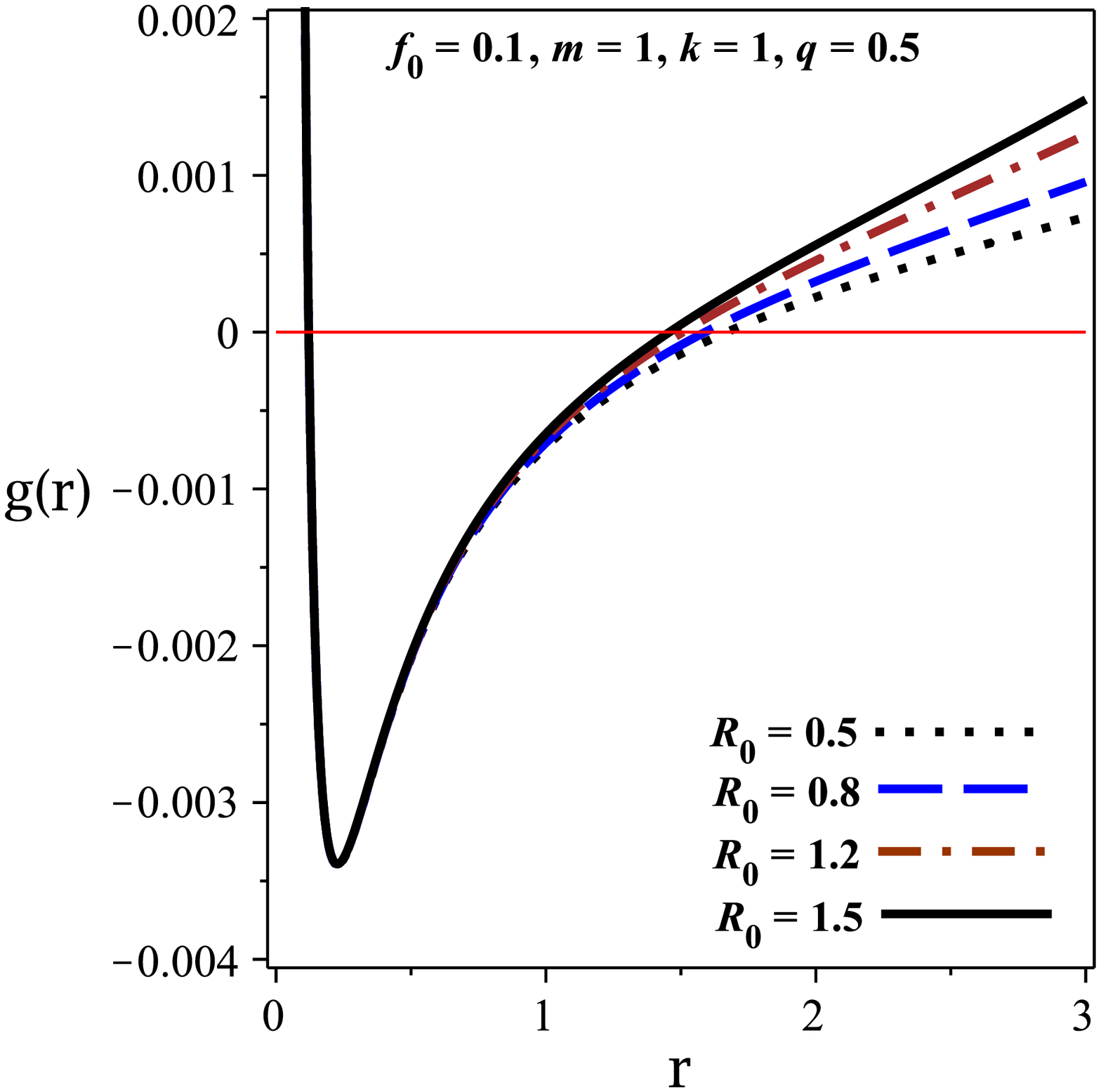} \includegraphics[width=0.4%
\linewidth]{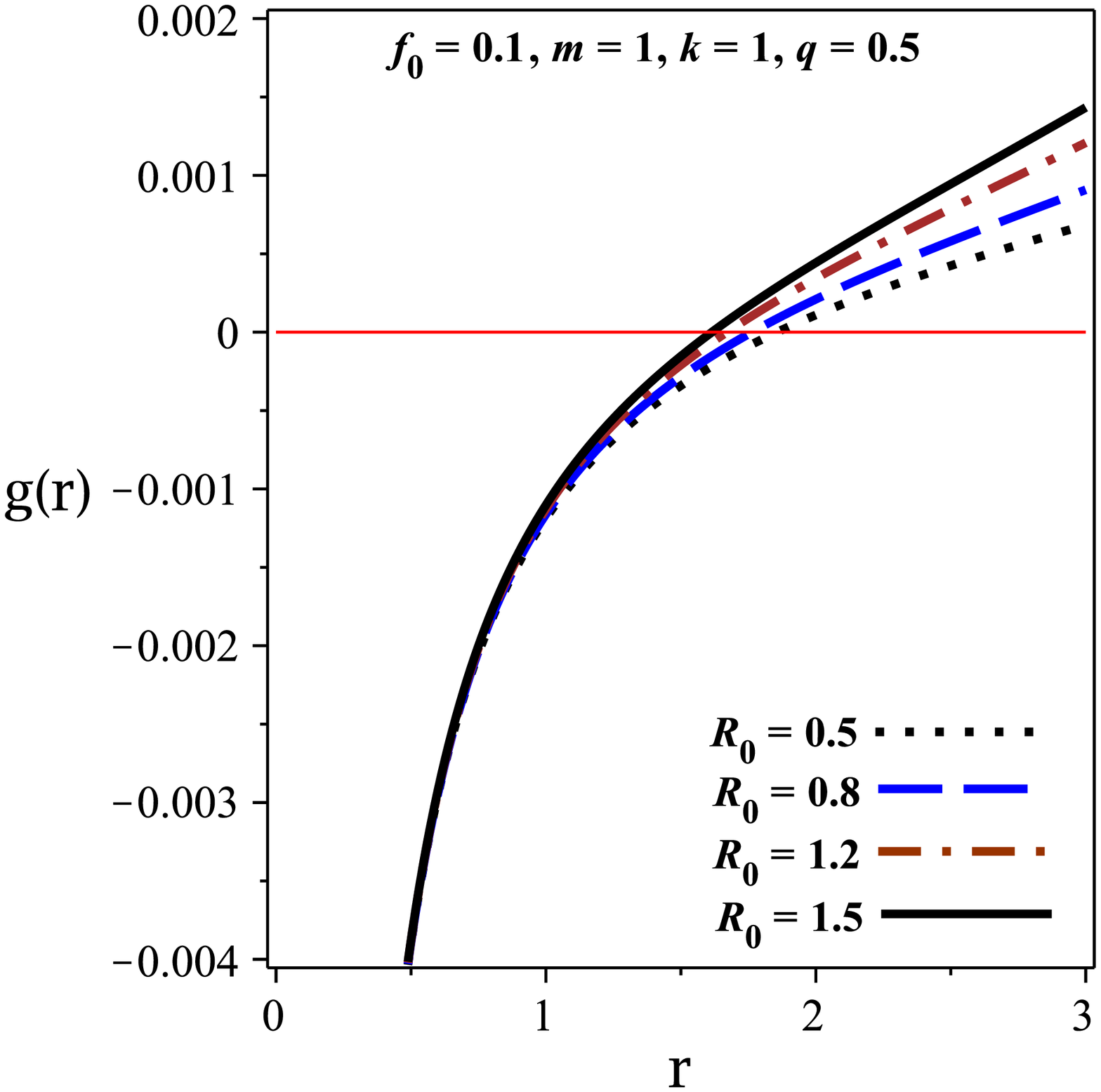}\newline
\caption{The metric function $g(r)$ versus $r$ for different values of the
parameters. Left panels for $\protect\eta =1$, and right panels for $\protect%
\eta =-1$}
\label{Fig1}
\end{figure}

To show that there is at least an event horizon in which covers the
singularity, we have to find the real roots of the obtained metric function (%
\ref{g(r)F(R)}). We plot the metric function versus $r$\ in Fig. \ref{Fig1}.
As shown in Fig. \ref{Fig1}, there is an event horizon for the obtained
metric function. Our findings indicate that the obtained solution in Eq. (%
\ref{g(r)F(R)}) is related to the black hole solution in $F(R)$ gravity with
Maxwell or phantom fields. It is worthwhile to mention that the solution (%
\ref{g(r)F(R)}) reduces to Reissner-Nordstr\"{o}m-(A)dS, when $f_{R_{0}}=1$, 
$R_{0}=-4\Lambda $ and $\eta =1$. In addition, we encounter with the
anti-Reissner-Nordstr\"{o}m-(A)dS (or phantom), when $f_{R_{0}}=1$, $%
R_{0}=-4\Lambda $ and $\eta =-1$. Also, we restrict ourselves to $%
f_{R_{0}}\neq -1$.

\section{Thermodynamics}

\label{sec2} 

Now, we are going to calculate the conserved and thermodynamic quantities of
the topological AdS phantom black hole solutions in $F(R)$ gravity to check
the first law of thermodynamics.

For studying the thermodynamic properties of the obtained black hole
solutions, it is necessary to express the mass ($m$) in terms of the radius
of the event horizon $r_{+}$ and the charge $q$ as follows. Equating $%
g_{tt}=g(r)$\ to zero, we have 
\begin{equation}
m=\frac{kr_{+}}{2}+\frac{R_{0}r_{+}^{3}}{24}+\frac{\eta q^{2}}{2\left(
1+f_{R_{0}}\right) r_{+}}.  \label{mm}
\end{equation}

Here, we want to obtain the Hawking temperature for these black holes. The
superficial gravity of a black hole is given by 
\begin{equation}
\kappa =\left. \frac{g_{tt}^{\prime }}{2\sqrt{-g_{tt}g_{rr}}}=\right\vert
_{r=r_{+}}=\left. \frac{g^{\prime }(r)}{2}\right\vert _{r=r_{+}},  \label{k}
\end{equation}%
where $r_{+}$ is the radius of the events horizon. Considering the obtained
metric function (\ref{g(r)F(R)}), and by substituting the mass (\ref{mm})\
within the equation (\ref{k}), one can calculate the superficial gravity as 
\begin{equation}
\kappa =\frac{k}{2r_{+}}+\frac{R_{0}r_{+}}{8}-\frac{\eta q^{2}}{2\left(
1+f_{R_{0}}\right) r_{+}^{3}},
\end{equation}%
and by using the Hawking temperature as $T=\frac{\kappa }{2\pi }$, we can
extract it in the following form 
\begin{equation}
T=\frac{k}{4\pi r_{+}}+\frac{R_{0}r_{+}}{16\pi }-\frac{\eta q^{2}}{4\pi
\left( 1+f_{R_{0}}\right) r_{+}^{3}}.  \label{TemF(R)CPMI}
\end{equation}

The electric charge of black hole per unit volume, $\mathcal{V}$, can be
obtained by using the Gauss law as 
\begin{equation}
Q=\frac{\widetilde{Q}}{\mathcal{V}}=\frac{F_{tr}}{4\pi }\int_{0}^{2\pi
}\int_{0}^{\pi }\sqrt{g_{k}}d\theta d\varphi =\frac{q}{4\pi }  \label{Q}
\end{equation}%
where $F_{tr}=\frac{q}{r^{2}}$, and for case $t=$ constant and $r=$constant,
the determinant of metric tensor $g_{k}$ is $r^{4}\det \left( d\Omega
_{k}^{2}\right) $ (i.e., $g_{k}=\det \left( g_{k}\right) =r^{4}\det \left(
d\Omega _{k}^{2}\right) $). It is worthwhile to mention that in the above
equation, we consider $\mathcal{V}=\int_{0}^{2\pi }\int_{0}^{\pi }\sqrt{\det
\left( d\Omega _{k}^{2}\right) }d\theta d\varphi $, where it is the area of
a unit volume of constant ($t$, $r$) space. Notable, $\mathcal{V}$ is $4\pi $
for $k=1$.

Considering $F_{\mu \nu }=\partial _{\mu }A_{\nu }-\partial _{\nu }A_{\mu }$%
, one can find the nonzero component of the gauge potential in which is $%
A_{t}=-\int F_{tr}dr$, and therefore the electric potential at the event
horizon ($U$) with respect to the reference ($r\rightarrow \infty $) is
given by%
\begin{equation}
U=-\int_{r_{+}}^{+\infty }F_{tr}dr=\frac{q}{r_{+}}.  \label{elcpoF(R)CPMI}
\end{equation}

In order to obtain the entropy of black holes in $F(R)=R+f(R)$\ theory, one
can use a modification of the area law which means the Noether charge method 
\cite{Cognola2005} 
\begin{equation}
S=\frac{A(1+f_{R})}{4},  \label{SFR}
\end{equation}%
where $A$\ is the horizon area and is defined 
\begin{equation}
A=\left. \int_{0}^{2\pi }\int_{0}^{\pi }\sqrt{g_{\theta \theta }g_{\varphi
\varphi }}\right\vert _{r=r_{+}}=\left. r^{2}\right\vert
_{r=r_{+}}=r_{+}^{2},  \label{A}
\end{equation}%
so, the entropy of topological phantom AdS black holes per unit volume, $%
\mathcal{V}$, in $F(R)$ gravity is given by replacing the horizon area (\ref%
{A}) within Eq. (\ref{SFR}) as

\begin{equation}
S=\frac{\widetilde{S}}{\mathcal{V}}=\frac{(1+f_{R})r_{+}^{2}}{4},  \label{S}
\end{equation}%
which indicates that the area law does not hold for the black hole solutions
in $R+f(R)$ gravity.

Using Ashtekar-Magnon-Das (AMD) approach \cite{AMDI,AMDII}, we find the
total mass of these black holes per unit volume, $\mathcal{V}$, in $F(R)$
gravity as 
\begin{equation}
M=\frac{\widetilde{M}}{\mathcal{V}}=\frac{m\left( 1+f_{R}\right) }{4\pi },
\label{AMDMass}
\end{equation}%
where substituting the mass (\ref{mm}) within the equation (\ref{AMDMass}),
yields 
\begin{equation}
M=\frac{\left( 1+f_{R}\right) r_{+}}{2}\left( k+\frac{R_{0}r_{+}^{2}}{12}%
\right) +\frac{\eta q^{2}}{2r_{+}}.  \label{MM}
\end{equation}

It is straightforward to show that the conserved and thermodynamics
quantities satisfy the first law of thermodynamics 
\begin{equation}
dM=TdS+\eta UdQ,
\end{equation}%
where $T=\left( \frac{\partial M}{\partial S}\right) _{Q}$, and $\eta
U=\left( \frac{\partial M}{\partial Q}\right) _{S}$, and they are in
agreement with those of calculated in Eqs. (\ref{TemF(R)CPMI}) and (\ref%
{elcpoF(R)CPMI}), respectively.


\section{Thermal Stability}

\label{sec3} 

Considering the black hole as a thermodynamic system, we want to study the
local and global stability. In the following, we investigate the effects of
the topological constant ($k$), the constant scalar curvature ($R_{0}$), and
the parameter of $\eta $ on the local and global stability of the
topological phantom (A)dS black holes in $F(R)$ gravity.


\subsection{Local Stability}


Here, we would like to study the local stability of the topological phantom
(A)dS black holes in the context of $F(R)$ gravity. For this purpose, by
considering these black holes we will study the heat capacity and the
geometrothemodynamics.


\subsubsection{Heat Capacity}


In the canonical ensemble context, a thermodynamic system's local stability
can be studied by heat capacity. The heat capacity carries crucial
information regarding the thermal structure of the black holes. This
quantity includes three specific exciting pieces of information:

i) The discontinuities of heat capacity mark the possible thermal phase
transitions that the system can undergo.

ii) The sign of it determines whether the system is thermally stable or not.
In other words, the positivity corresponds to thermal stability while the
opposite indicates instability.

iii) The roots of heat capacity are also of interest since they may yield
the possible changes between stable/unstable states or bound points.

Due to these important points, we want to calculate the heat capacity of the
solutions and investigation of local stability of the black holes by using
such quantity.

Before obtaining the heat capacity, let us first re-write the total mass of
the black hole (\ref{MM}) in terms of the entropy (\ref{S}) in the following
form 
\begin{equation}
M\left( S,Q\right) =\frac{\pi ^{2}Q^{2}\left( 1+f_{R}\right) \eta +\frac{%
S\left( SR_{0}+3\left( 1+f_{R}\right) k\right) }{12}}{\pi \sqrt{S\left(
1+f_{R}\right) }},  \label{MSQ}
\end{equation}%
using the equation (\ref{MSQ}), we re-write the temperature in the following
form 
\begin{equation}
T=\left( \frac{\partial M\left( S,Q\right) }{\partial S}\right) _{Q}=\frac{%
\frac{S\left( SR_{0}+\left( 1+f_{R}\right) k\right) }{4}-\pi ^{2}Q^{2}\left(
1+f_{R}\right) \eta }{2\pi S^{3/2}\sqrt{1+f_{R}}}.  \label{TM}
\end{equation}

The heat capacity is defind as 
\begin{equation}
C_{Q}=\frac{T}{\left( \frac{\partial T}{\partial S}\right) _{Q}}=\frac{%
\left( \frac{\partial M\left( S,Q\right) }{\partial \widetilde{S}}\right)
_{Q}}{\left( \frac{\partial ^{2}M\left( S,Q\right) }{\partial S^{2}}\right)
_{Q}},  \label{Heat}
\end{equation}%
by considering Eqs. (\ref{MSQ}) and (\ref{TM}), we can obtain the heat
capacity in form 
\begin{equation}
C_{Q}=\frac{2S^{2}\left( \frac{SR_{0}}{\left( 1+f_{R}\right) }+k\right)
-8\pi ^{2}Q^{2}S\eta }{12\pi ^{2}Q^{2}\eta +S\left( \frac{SR_{0}}{\left(
1+f_{R}\right) }-k\right) }.  \label{Heat1}
\end{equation}

In the context of black holes, it is argued that the root of heat capacity $%
\left( C_{Q}=T=0\right) $ is representing a border line between physical $%
\left( T>0\right) $\ and non-physical $\left( T<0\right) $ black holes. We
call it a physical limitation point. Indeed, the system in the case of this
physical limitation point has a change in sign of the heat capacity. Also,
it is believed that the divergences of the heat capacity represent phase
transition critical points of black holes. So, the phase transition critical
and limitation points of the black holes in the context of the heat capacity
are calculated with the following relations 
\begin{equation}
\left\{ 
\begin{array}{ccc}
T=\left( \frac{\partial M\left( S,Q\right) }{\partial S}\right) _{Q}=0, &  & 
\text{physical limitation points} \\ 
&  &  \\ 
\left( \frac{\partial ^{2}M\left( S,Q\right) }{\partial S^{2}}\right) _{Q}=0
&  & \text{phase transition critical points}%
\end{array}%
\right. .  \label{PhysBound}
\end{equation}

Using Eq. (\ref{TM}) and solving it in terms of the entropy, we can obtain
physical limitation points as 
\begin{equation}
\left\{ 
\begin{array}{c}
S_{root_{1}}=\frac{-\left( 1+f_{R}\right) k}{2R_{0}}+\frac{\sqrt{\left[
\left( 1+f_{R}\right) k^{2}+16\pi ^{2}Q^{2}\eta R_{0}\right] \left(
1+f_{R}\right) }}{2R_{0}} \\ 
\\ 
S_{root_{2}}=\frac{-\left( 1+f_{R}\right) k}{2R_{0}}-\frac{\sqrt{\left[
\left( 1+f_{R}\right) k^{2}+16\pi ^{2}Q^{2}\eta R_{0}\right] \left(
1+f_{R}\right) }}{2R_{0}}%
\end{array}%
\right. .
\end{equation}

To have the real root(s), we have to respect $\left( 1+f_{R}\right)
k^{2}+16\pi ^{2}Q^{2}\eta R_{0}\geq 0$. This constraint gives us information
about the effects of different parameters on the roots of temperature (\ref%
{TM}). For example, the temperature has one root for $k=0$, provided $\eta
R_{0}>0$. For $k=\pm 1$, the temperature has two roots when $\eta R_{0}<0$,
provided $R_{0}>\frac{-\left( 1+f_{R}\right) k^{2}}{6\pi ^{2}Q^{2}\eta }$.
In addition, the relation $R_{0}>\frac{-\left( 1+f_{R}\right) k^{2}}{6\pi
^{2}Q^{2}\eta }$, imposes that for a large value of the electrical charge
and $k=\pm 1$, the temperature does not have any root when the constant
scalar curvature is negative. Indeed, the temperature of higher-charged
black holes does not have any root when $k=\pm 1$ and $R_{0}<0$.

In order to study the phase transition critical points (or divergence points
of the heat capacity), we have to solve the relation $\left( \frac{\partial
^{2}M\left( S,Q\right) }{\partial S^{2}}\right) _{Q}=0$. So, we have 
\begin{equation}
\left\{ 
\begin{array}{c}
S_{div_{1}}=\frac{\left( 1+f_{R}\right) k}{2R_{0}}-\frac{\sqrt{\left[ \left(
1+f_{R}\right) k^{2}-48\pi ^{2}Q^{2}\eta R_{0}\right] \left( 1+f_{R}\right) }%
}{2R_{0}} \\ 
\\ 
S_{div_{2}}=\frac{\left( 1+f_{R}\right) k}{2R_{0}}+\frac{\sqrt{\left[ \left(
1+f_{R}\right) k^{2}-48\pi ^{2}Q^{2}\eta R_{0}\right] \left( 1+f_{R}\right) }%
}{2R_{0}}%
\end{array}%
\right. ,  \label{rdivHeat}
\end{equation}%
where indicate that we have to respect $\left( 1+f_{R}\right) k^{2}-48\pi
^{2}Q^{2}\eta R_{0}\geq 0$, for having the real divergent point(s). Our
analysis shows that the heat capacity has one divergent point for $k=0$,
provided $\eta R_{0}<0$, and also there is no divergent point when $\eta
R_{0}>0$. According to Eq. (\ref{rdivHeat}), for large values of the
electrical charge, the heat capacity does not have any divergent point. In
other words, the heat capacity of the higher-charged black holes does not
have any divergent point. For $k=+1$ ( $k=-1$), the heat capacity has two
divergence points when $\eta >0$ ($\eta <0$)\ and $R_{0}>0$ ($R_{0}<0$).

\begin{figure}[tbph]
\centering
\includegraphics[width=0.4\linewidth]{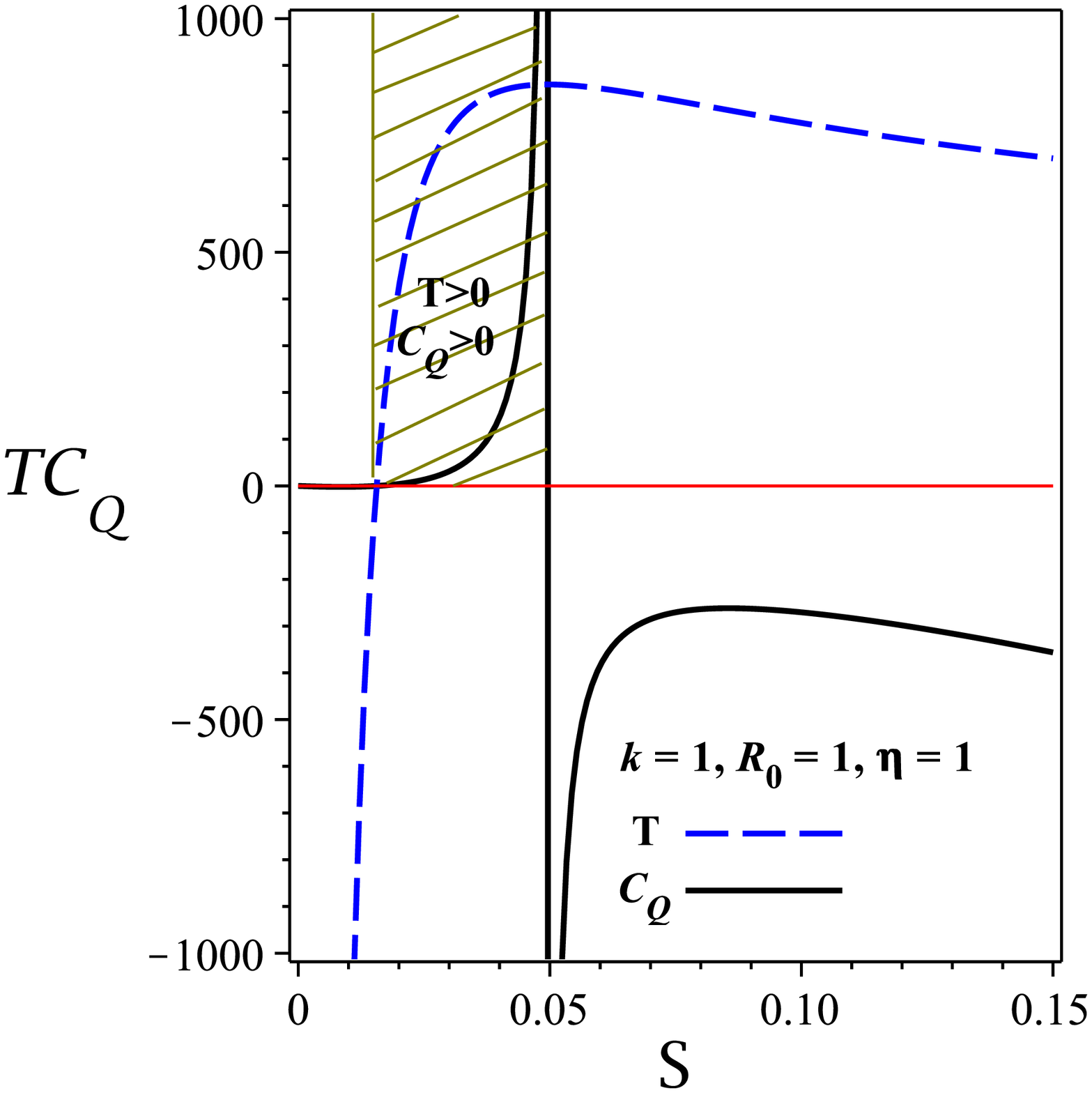} \includegraphics[width=0.4%
\linewidth]{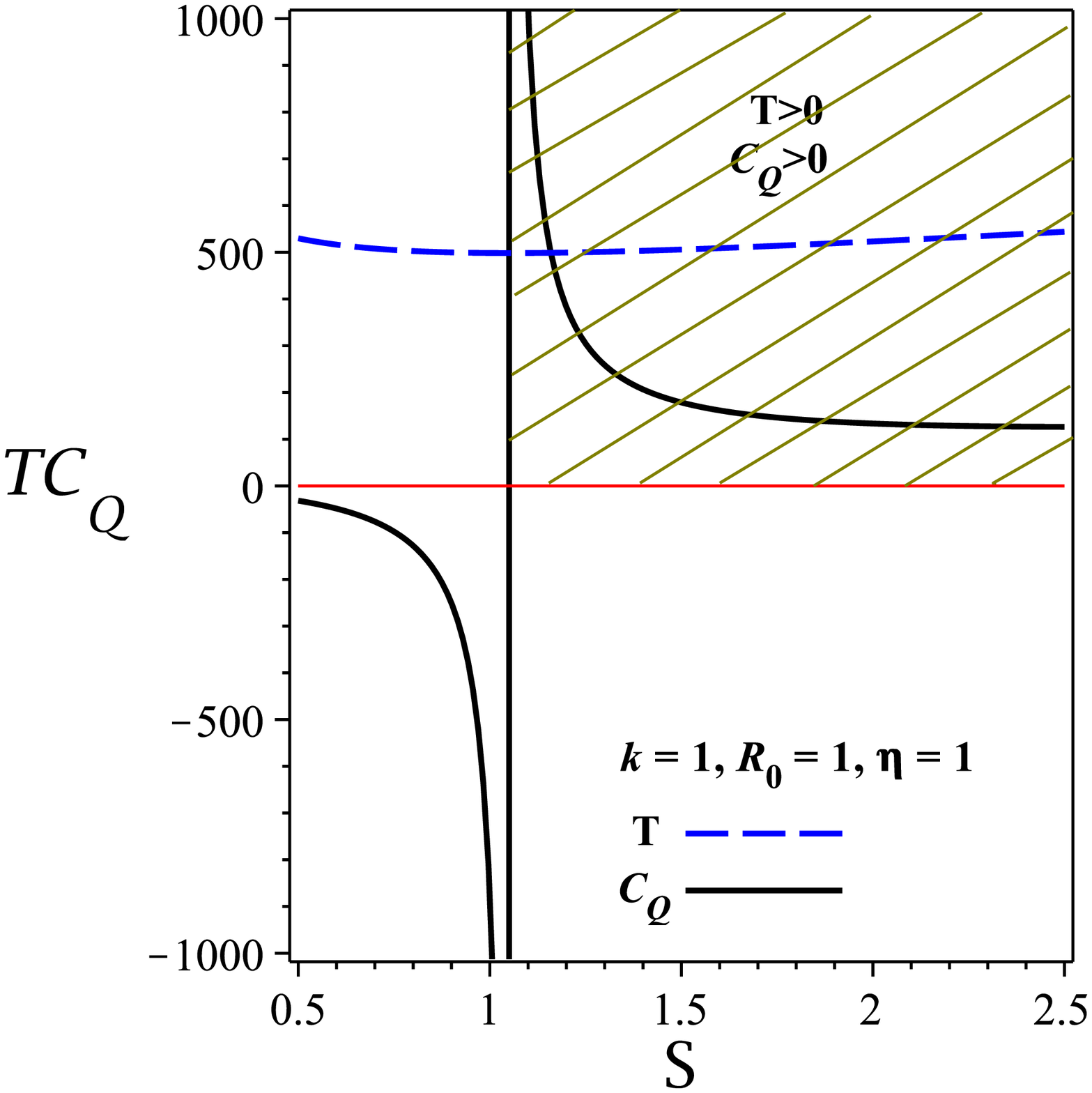}\newline
\includegraphics[width=0.4\linewidth]{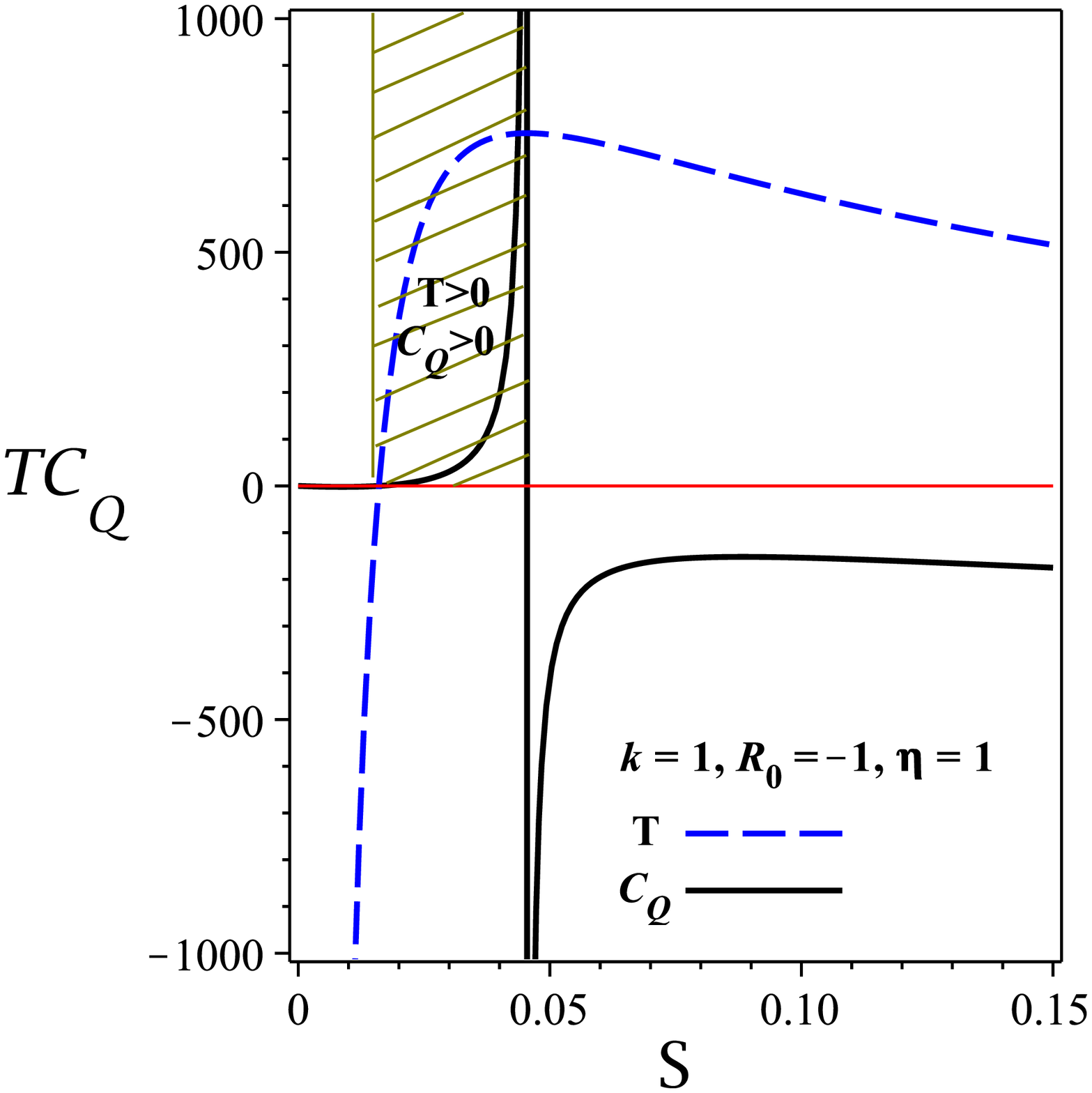} \includegraphics[width=0.4%
\linewidth]{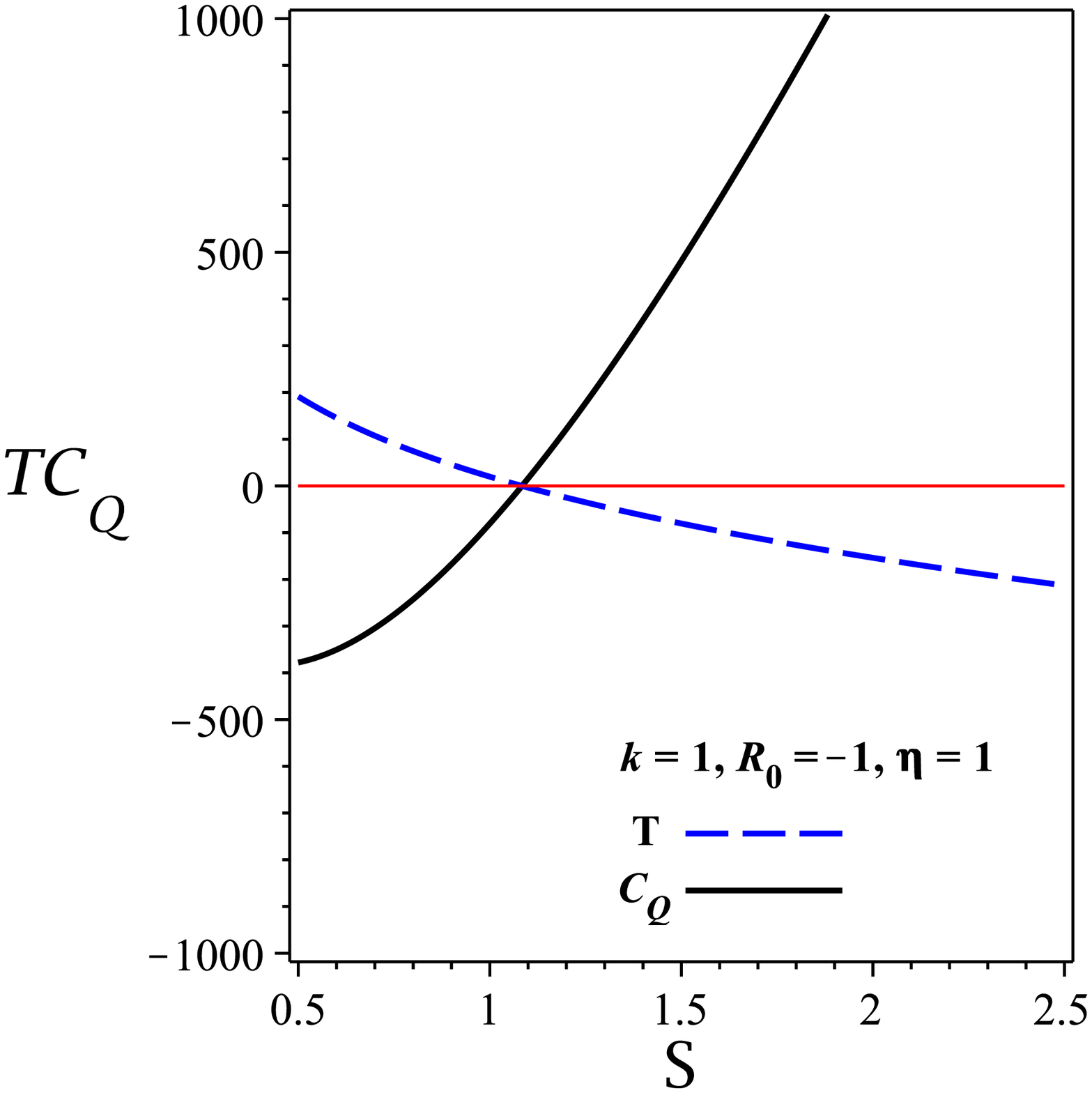}\newline
\includegraphics[width=0.4\linewidth]{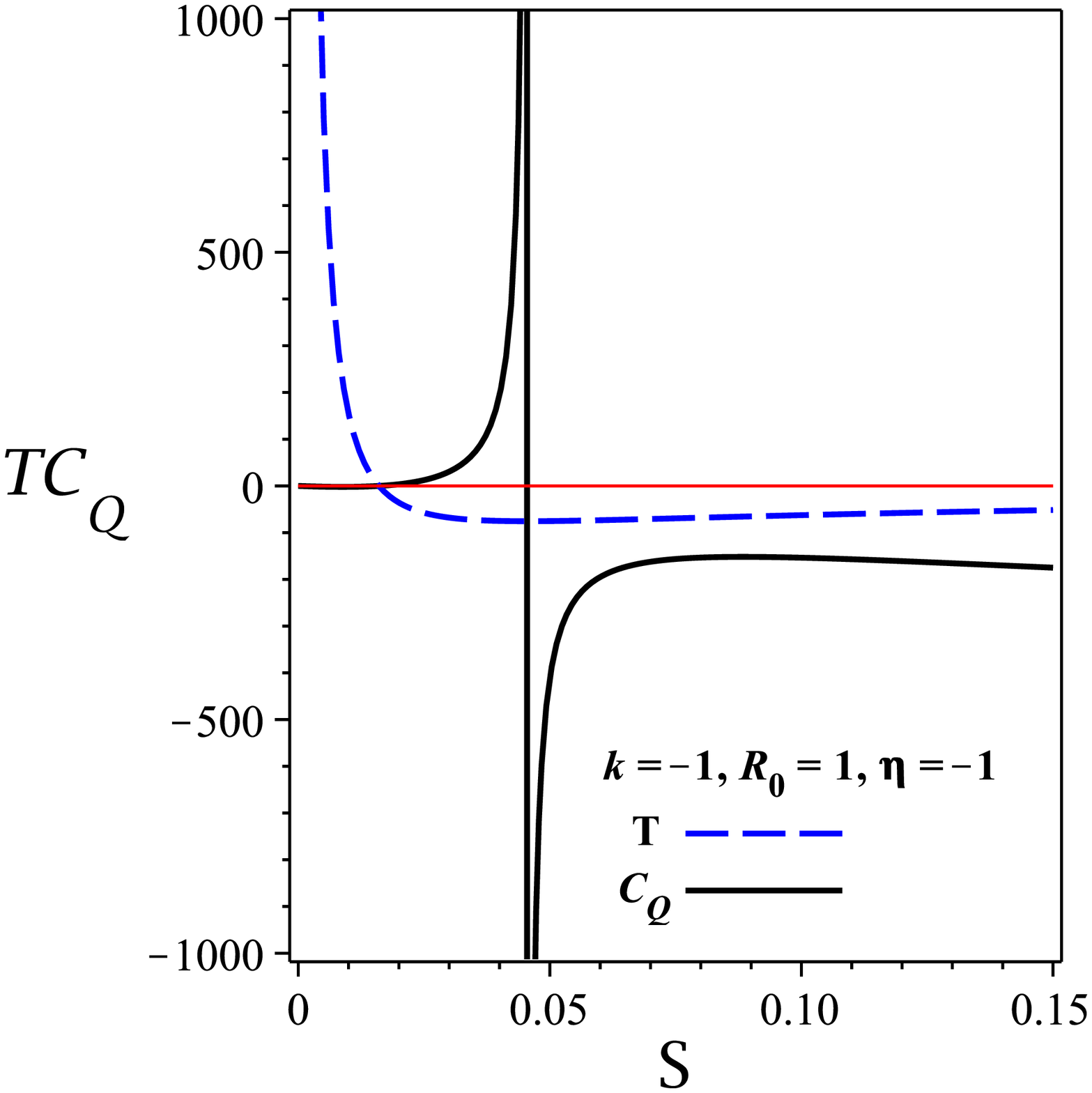} \includegraphics[width=0.4%
\linewidth]{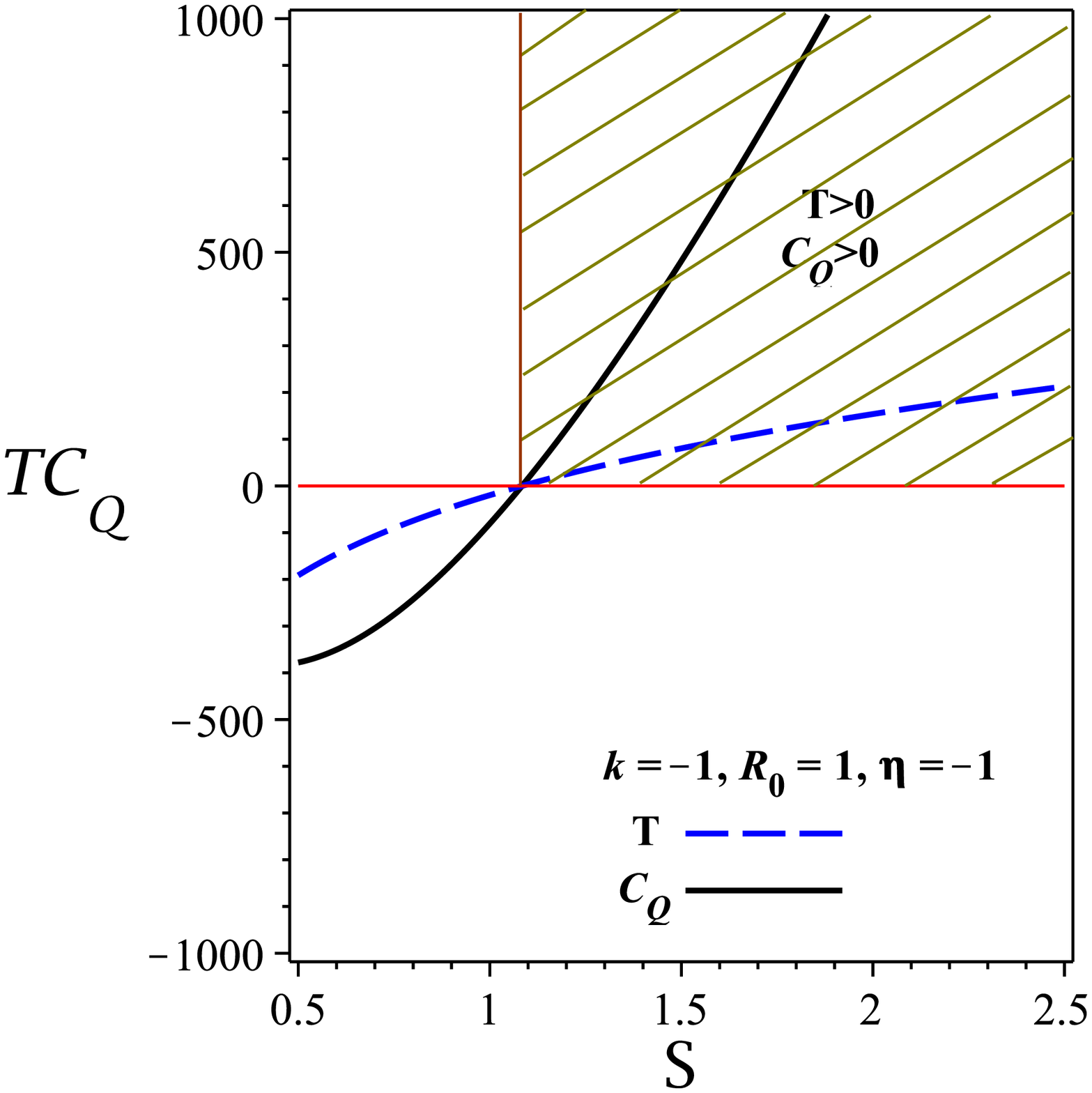}\newline
\includegraphics[width=0.4\linewidth]{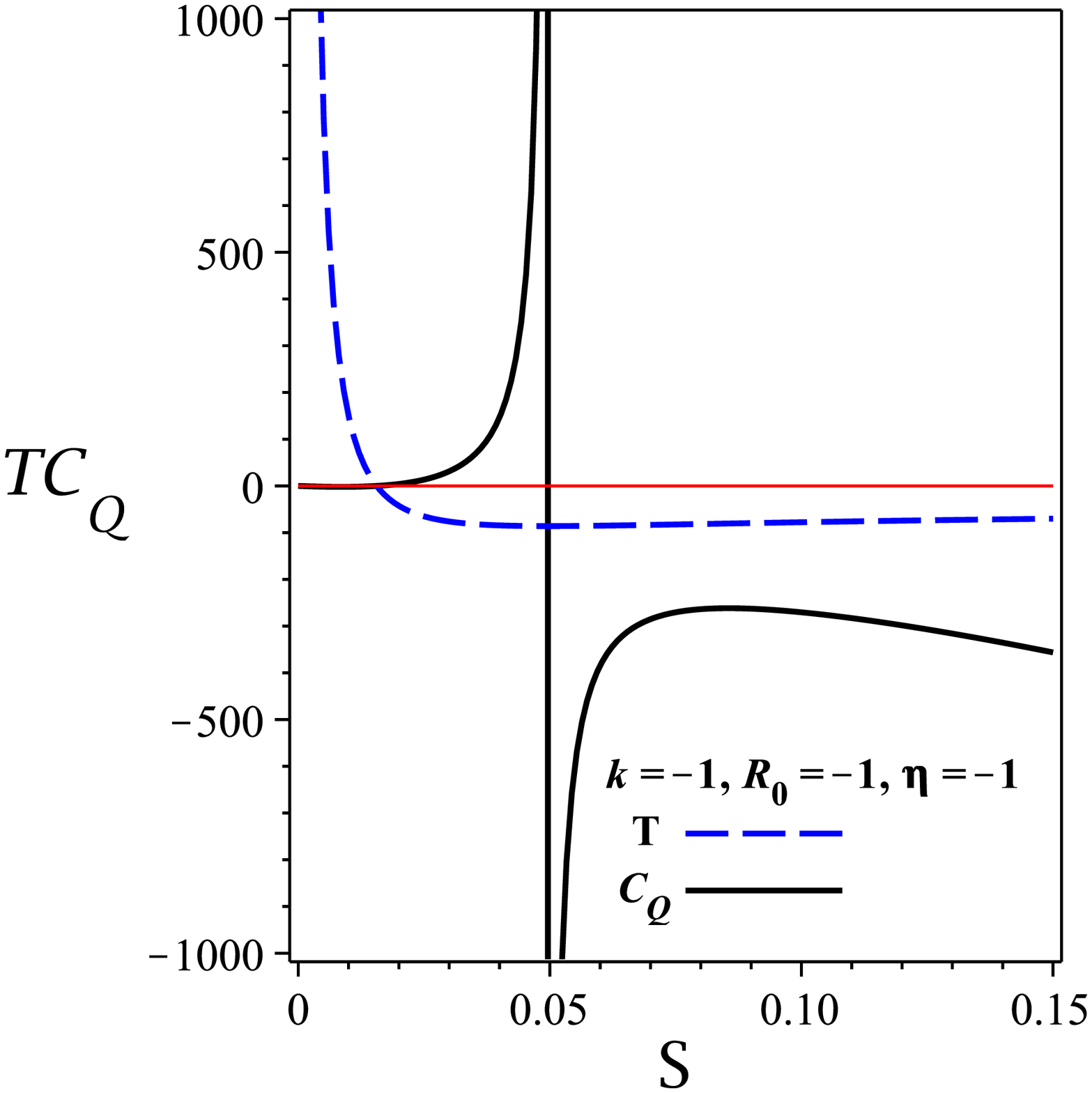} \includegraphics[width=0.4%
\linewidth]{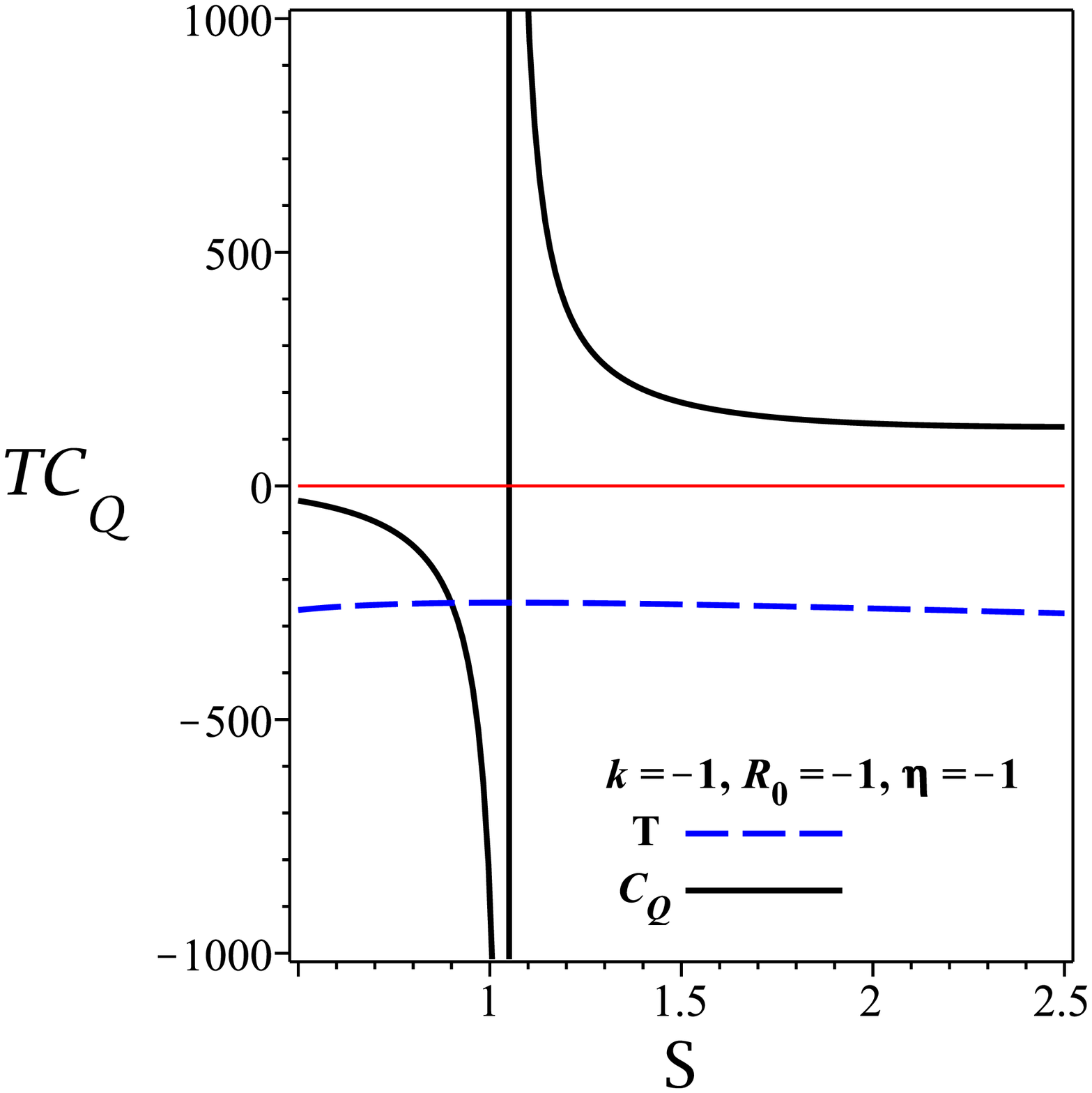}\newline
\caption{The heat capacity ($C_{Q}$) and temperature ($T$) versus $S$ for $%
Q=0.02$ and $f_{R}=0.1$. We plot them in different scales in order to be
more clear.}
\label{Fig2}
\end{figure}

Now we can evaluate the local stability by using the behavior of temperature
and heat capacity. For this purpose, we plot Fig. \ref{Fig2} and analyze
them with more details in Table. \ref{tab1}.

\begin{table*}[tbp]
\caption{The local stability of the black holes for $Q=0.02$, $f_{R}=0.1$,
and different values of $k$. First three rows are for $R_{0}=1$, and $%
\protect\eta =1$. Second three rows are for $R_{0}=-1$, and $\protect\eta =1$%
. Third three rows are for $R_{0}=1$, and $\protect\eta =-1$. Fourth three
rows are for $R_{0}=-1$, and $\protect\eta =-1$.}
\label{tab1}
\begin{center}
\begin{tabular}{|c|c|c|c|c|c|}
\hline
$k$ & number of $S_{root}$ & number of $S_{div}$ & physical area ($T>0$) & $%
C_{Q}>0$ & local stability and physical area \\ \hline\hline
$+1$ & 1 & 2 & $S>S_{root}$ & $%
\begin{array}{c}
S_{root}<S<S_{div_{1}} \\ 
S>S_{div_{2}}%
\end{array}
$ & $%
\begin{array}{c}
S_{root}<S<S_{div_{1}} \\ 
S>S_{div_{2}}%
\end{array}
$ \\ \hline
$0$ & 1 & 0 & $S>S_{root}$ & $S>S_{root}$ & $S>S_{root}$ \\ \hline
$-1$ & 1 & 0 & $S>S_{root}$ & $S>S_{root}$ & $S>S_{root}$ \\ \hline\hline
$+1$ & 2 & 1 & $S_{root_{1}}<S<S_{root_{2}}$ & $%
\begin{array}{c}
S_{root_{1}}<S<S_{div} \\ 
S>S_{root_{2}}%
\end{array}
$ & $S_{root_{1}}<S<S_{div}$ \\ \hline
$0$ & 0 & 1 & no area & $S>S_{div}$ & no area \\ \hline
$-1$ & 0 & 1 & no area & $S>S_{div}$ & no area \\ \hline\hline
$+1$ & 0 & 1 & always positive & $S>S_{div}$ & $S>S_{div}$ \\ \hline
$0$ & 0 & 1 & always positive & $S>S_{div}$ & $S>S_{div}$ \\ \hline
$-1$ & 2 & 1 & $%
\begin{array}{c}
S<S_{root_{1}} \\ 
S>S_{root_{2}}%
\end{array}
$ & $%
\begin{array}{c}
S_{root_{1}}<S<S_{div} \\ 
S>S_{root_{2}}%
\end{array}
$ & $S>S_{root_{2}}$ \\ \hline\hline
$+1$ & 1 & 0 & $S<S_{root}$ & $S>S_{root}$ & no area \\ \hline
$0$ & 1 & 0 & $S<S_{root}$ & $S>S_{root}$ & no area \\ \hline
$-1$ & 1 & 2 & $S<S_{root}$ & $%
\begin{array}{c}
S_{root}<S<S_{div_{1}} \\ 
S>S_{div_{2}}%
\end{array}
$ & no area \\ \hline
\end{tabular}%
\end{center}
\end{table*}

Our findings reveal some interesting behaviors which are:

i) The topological charged AdS black holes in $F(R)$ gravity satisfy the
local stability condition when they have large radii (or large entropy), see
the first three rows of Table. \ref{tab1}, for more details.

ii) The charged dS black holes with medium radii can be only locally stable
for $k=1$ (see the second three rows of Table. \ref{tab1}, for more details).

iii) The phantom AdS large black holes have local stability (see the third
three rows of Table. \ref{tab1}, for more details).

iv) There is no local stability area for the phantom AdS black holes with
different topological constants (see the fourth three rows in Table. \ref%
{tab1}, for more details).

We plotted the figure. \ref{Fig2}, for more details. Two up panels in Fig. %
\ref{Fig2} belong to the charged (A)dS black holes in $F(R)$ gravity for $%
k=+1$ and $k=-1$. Also, the two down panels are related to phantom (A)dS
black holes in $F(R)$ gravity for $k=+1$ and $k=-1$. In Fig. \ref{Fig2}, the
hatched areas belong to the physical and local stability of these black
holes.

\subsubsection{Geometrothermodynamics}


Here, we want to study the phase transition of the topological phantom (A)dS
black holes in $F(R)$ gravity through geometrothemodynamics. In the
geometrothemodynamics method, a thermodynamical metric (thermodynamical
phase space) is constructed by considering one of the thermodynamical
quantities as thermodynamical potential and other quantities as extensive
parameters. By calculating the Ricci scalar of such a thermodynamical metric
and determining its divergence point(s), one can obtain the phase transition
point(s) of the system. In this regard, several thermodynamical metrics have
been introduced in order to build a geometrical phase space by
thermodynamical quantities. The famous ones are the Weinhold \cite%
{Weinhold1,Weinhold2} , Ruppeiner \cite{Ruppeiner1,Ruppeiner2}, and Quevedo
metrics \cite{Quevedo1,Quevedo2}. It was previously argued that these
metrics may not provide us with a completely flawless mechanism for
evaluating the geometrothemodynamics of specific types of black holes (see
Refs. \cite{HPEM1,HPEM2,HPEM3,HPEM4,HPEM5,HPEM6}, for more details).
Recently, a new metric (which is known as HPEM metric \cite{HPEM1}), was
introduced in order to solve the problems that other metrics may confront
with it. In this section, we would like to investigate the phase transition
of the topological phantom (A)dS black holes in the non-extended phase space
via the geometrothemodynamics method which is described by the HPEM metric.

The HPEM metric is given by \cite{HPEM1} 
\begin{equation}
ds^{2}=\frac{SM_{S}}{M_{QQ}^{3}}\left( -M_{SS}dS^{2}+M_{QQ}dQ^{2}\right) ,
\end{equation}%
where $M_{S}=\frac{\partial M}{\partial S}$, $M_{SS}=\frac{\partial ^{2}M}{%
\partial S\partial S}$, and $M_{QQ}=\frac{\partial ^{2}M}{\partial Q\partial
Q}$. Since we are looking for the divergence points of the HPEM's Ricci
scalar and because its numerator is a smooth finite function, we focus on
the denominator of the HPEM's Ricci scalar. The denominator of the HPEM's
Ricci scalar is given by \cite{HPEM1} 
\begin{equation}
denom(R)=2S^{3}M_{S}^{3}M_{SS}^{2}.  \label{RR}
\end{equation}

\begin{figure}[tbph]
\centering
\includegraphics[width=0.4\linewidth]{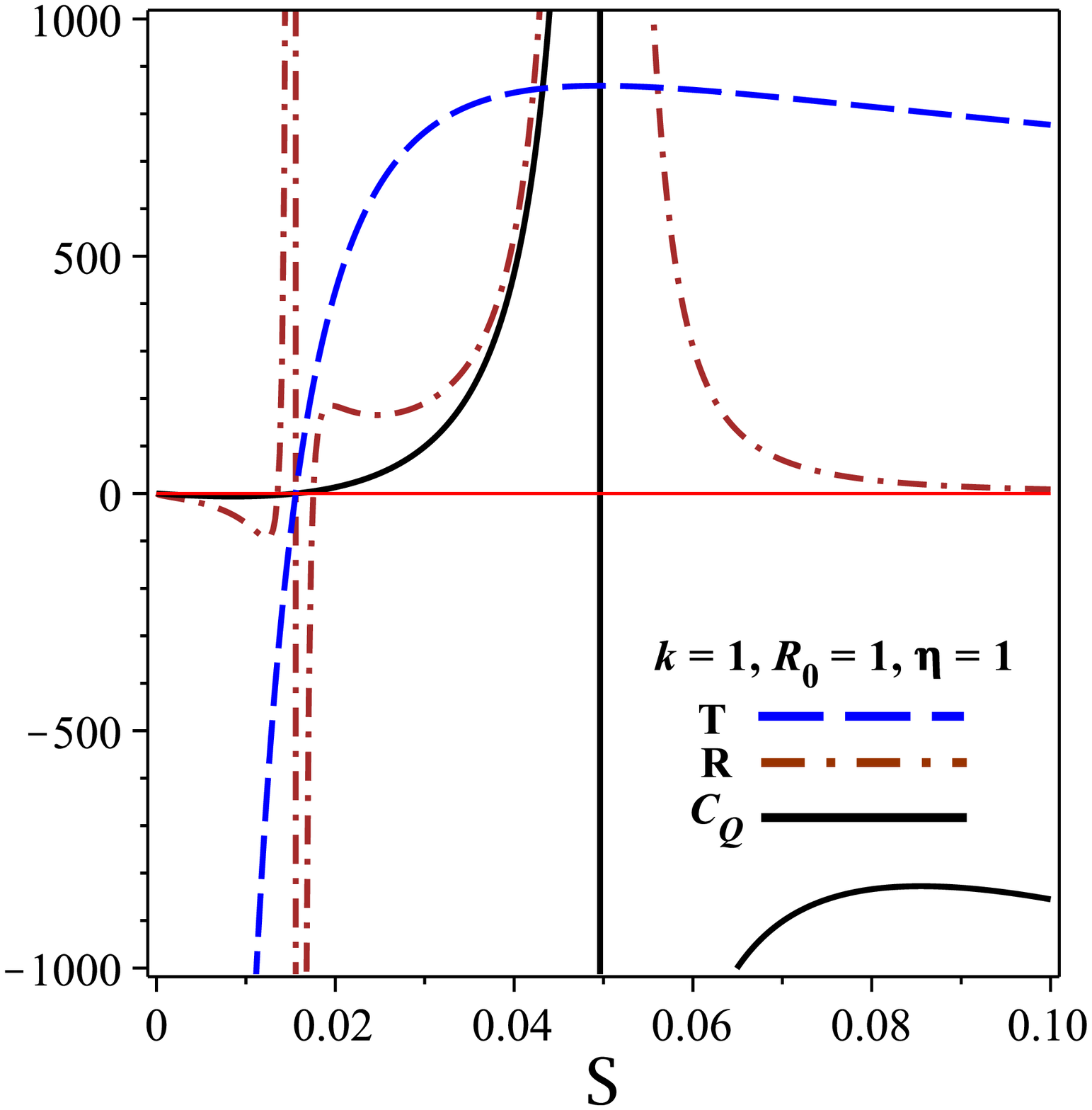} \includegraphics[width=0.4%
\linewidth]{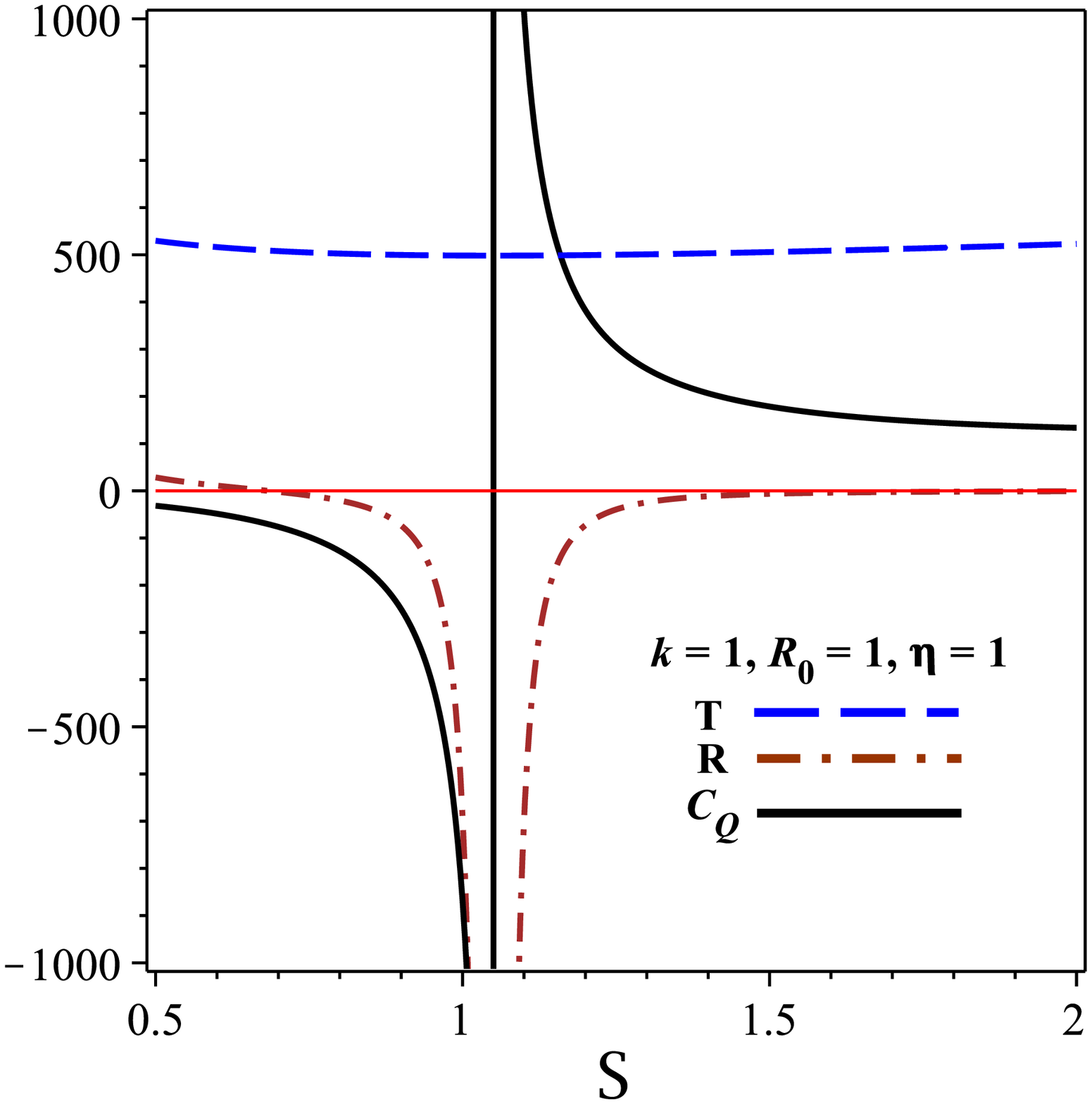}\newline
\includegraphics[width=0.4\linewidth]{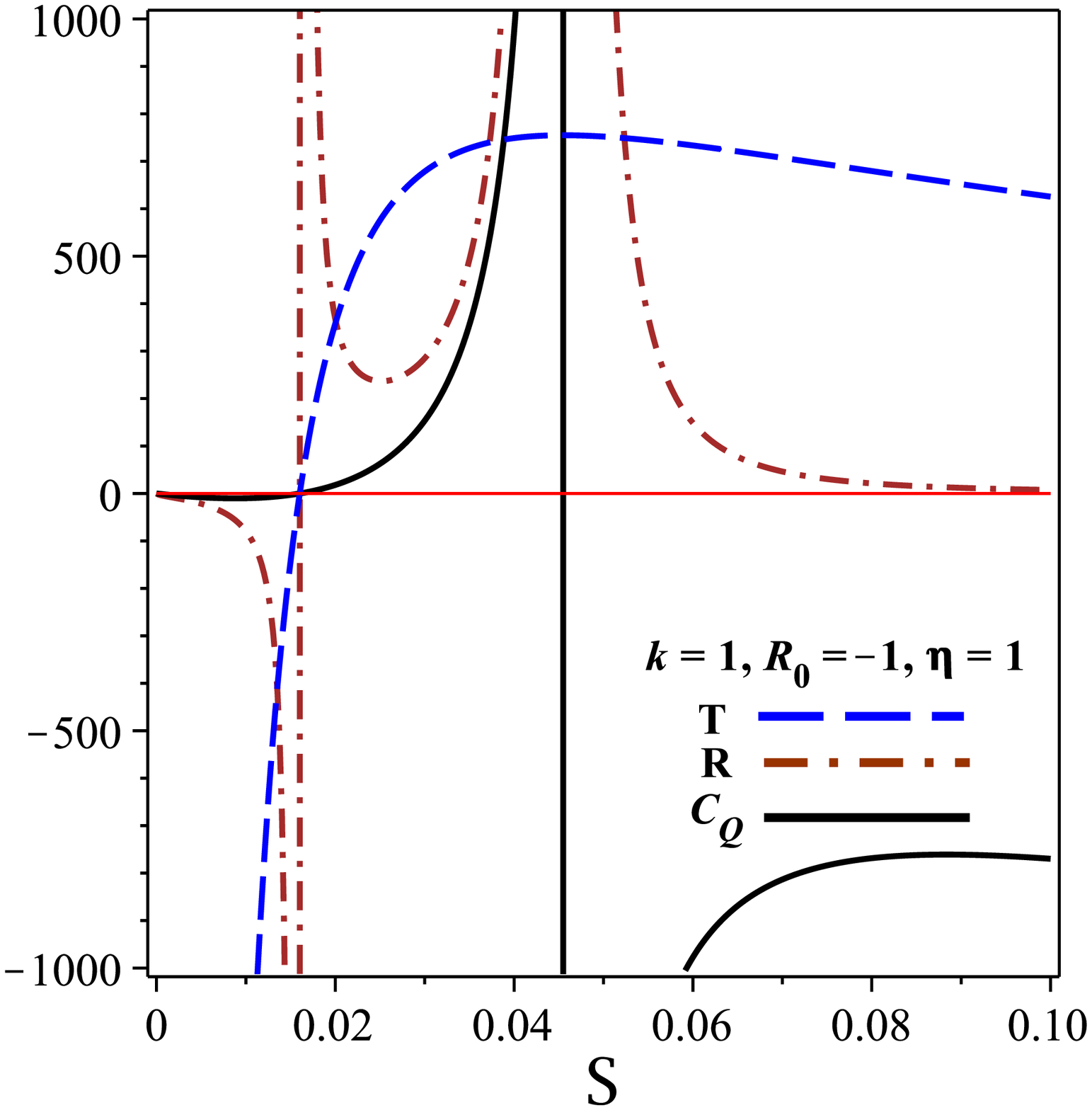} \includegraphics[width=0.4%
\linewidth]{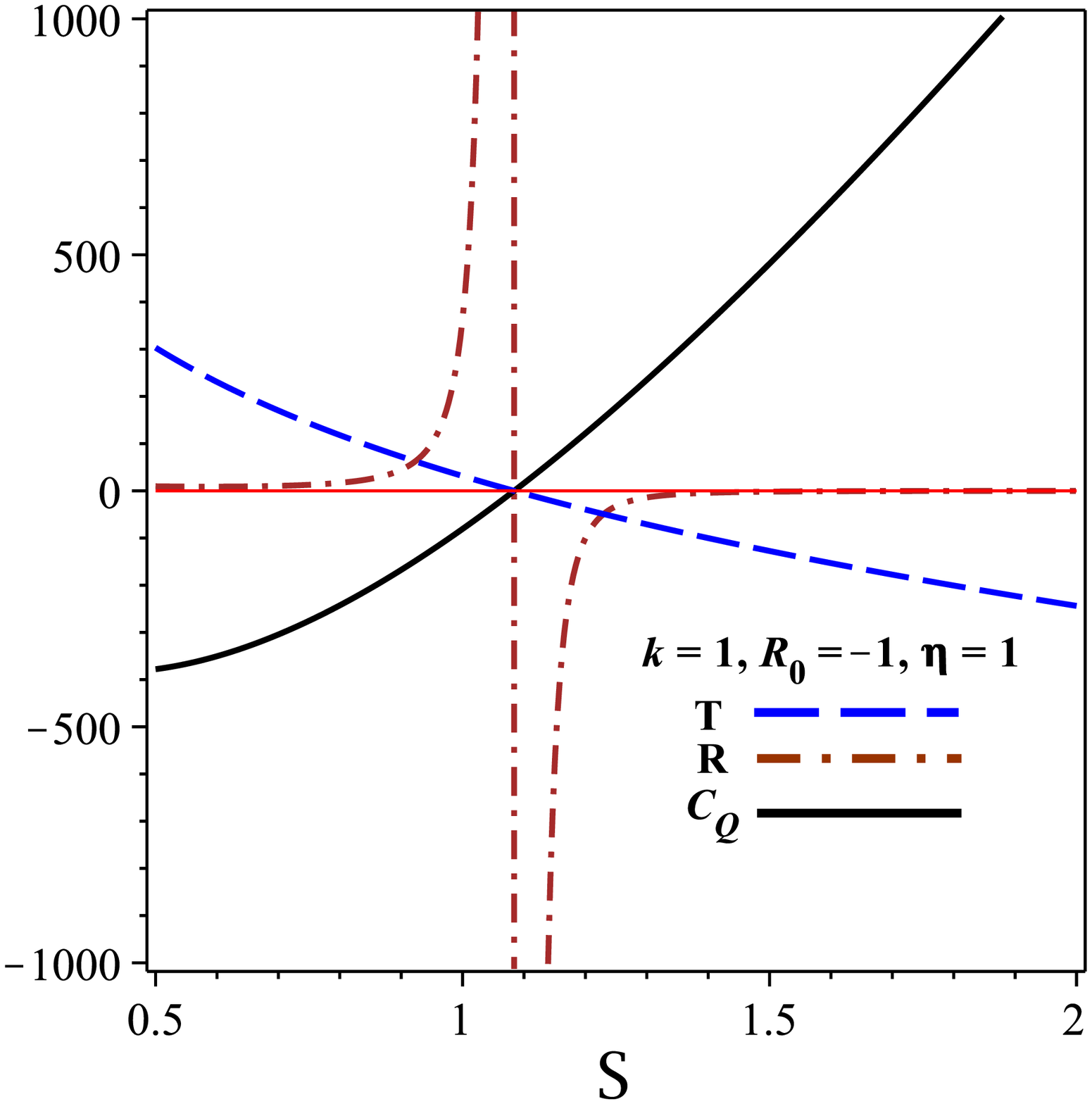}\newline
\includegraphics[width=0.4\linewidth]{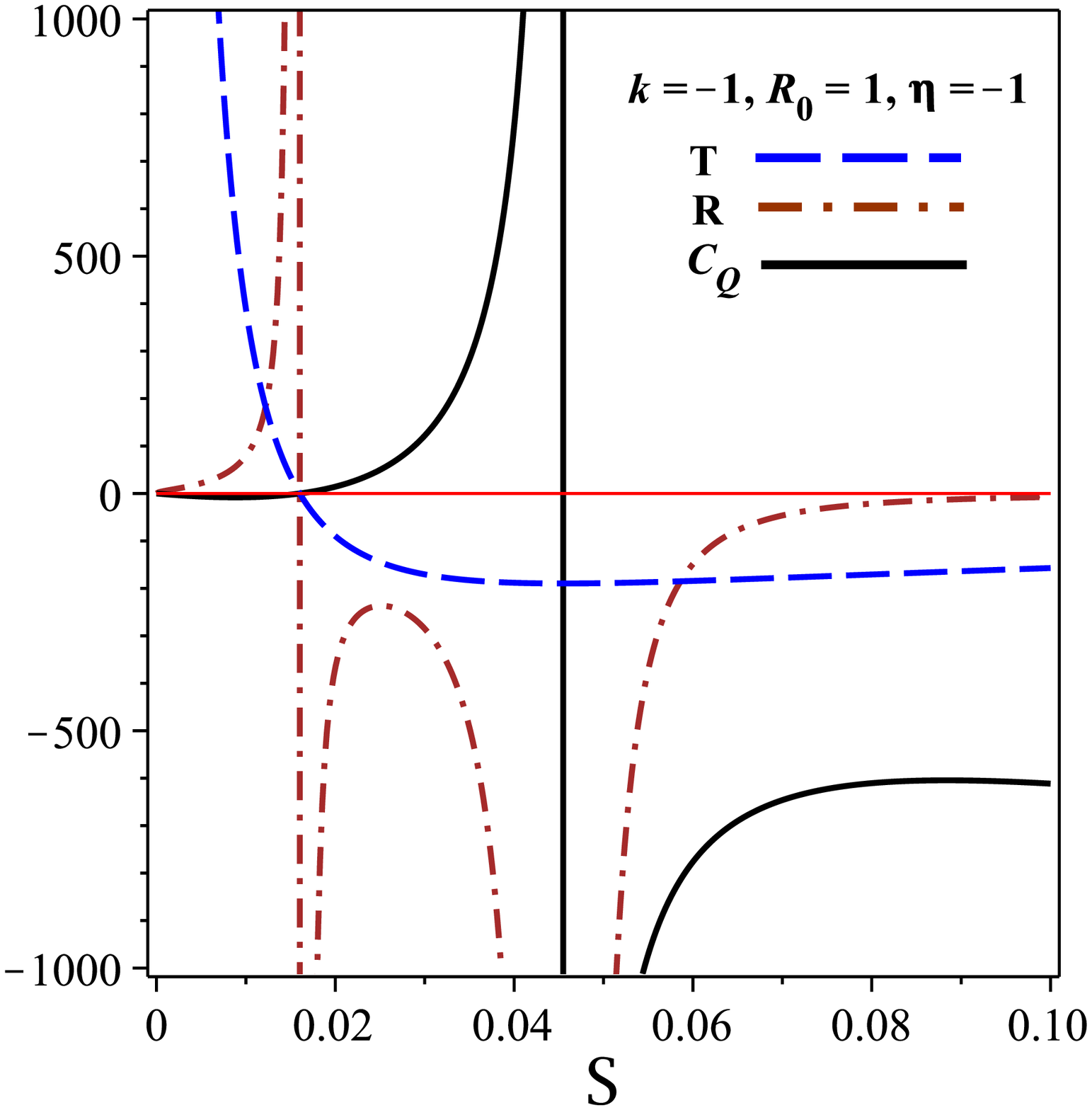} \includegraphics[width=0.4%
\linewidth]{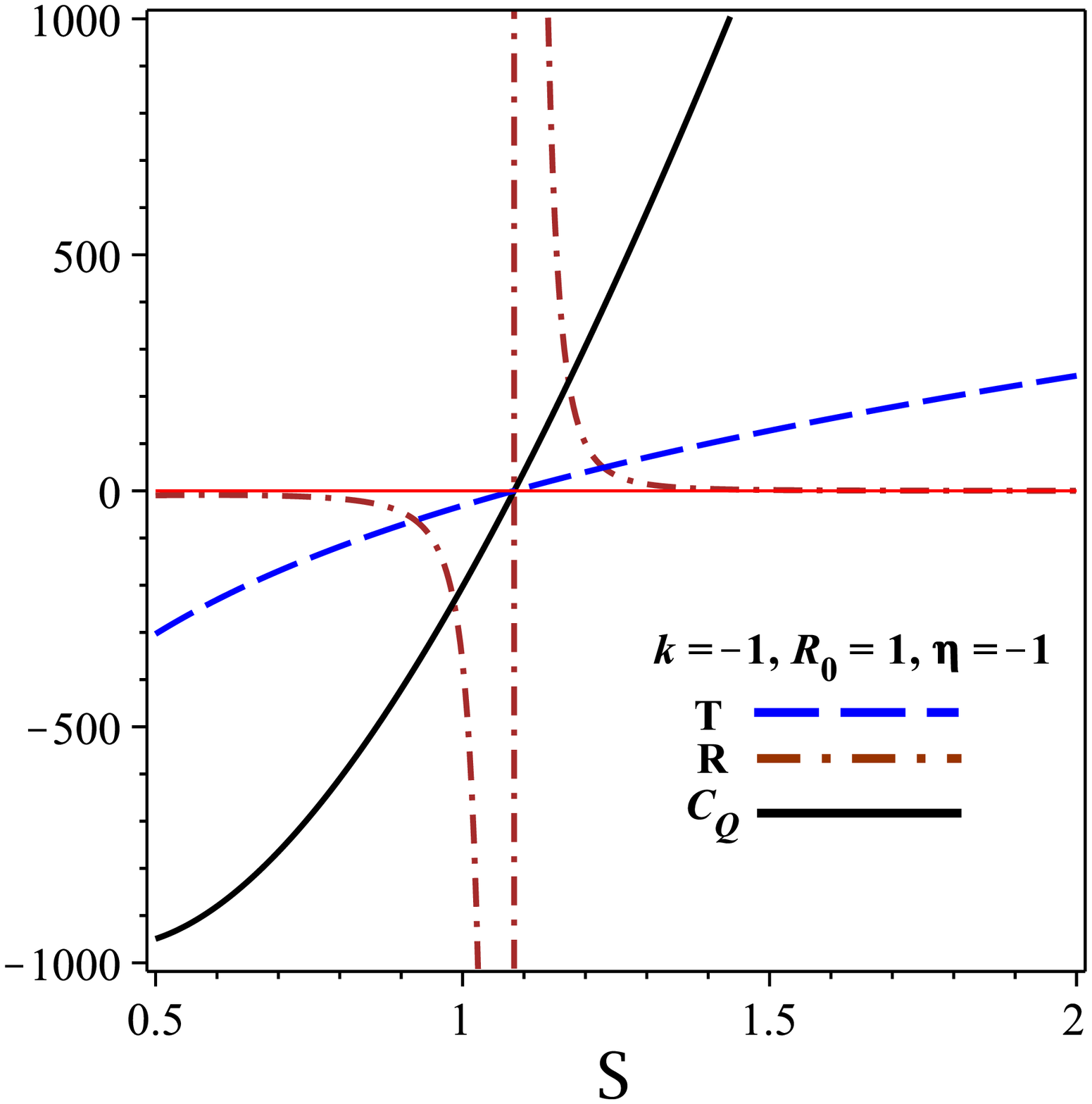}\newline
\includegraphics[width=0.4\linewidth]{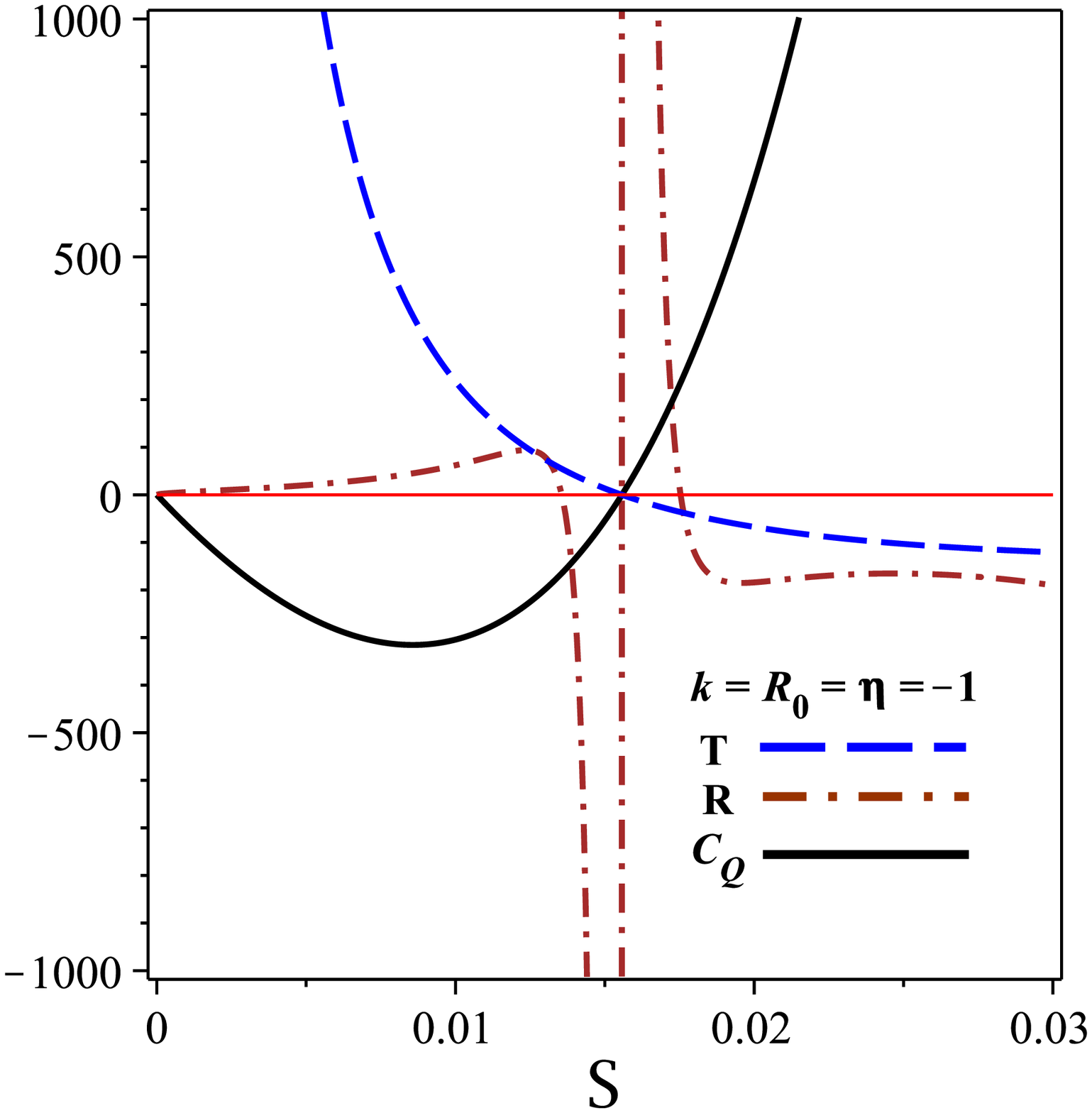} \includegraphics[width=0.4%
\linewidth]{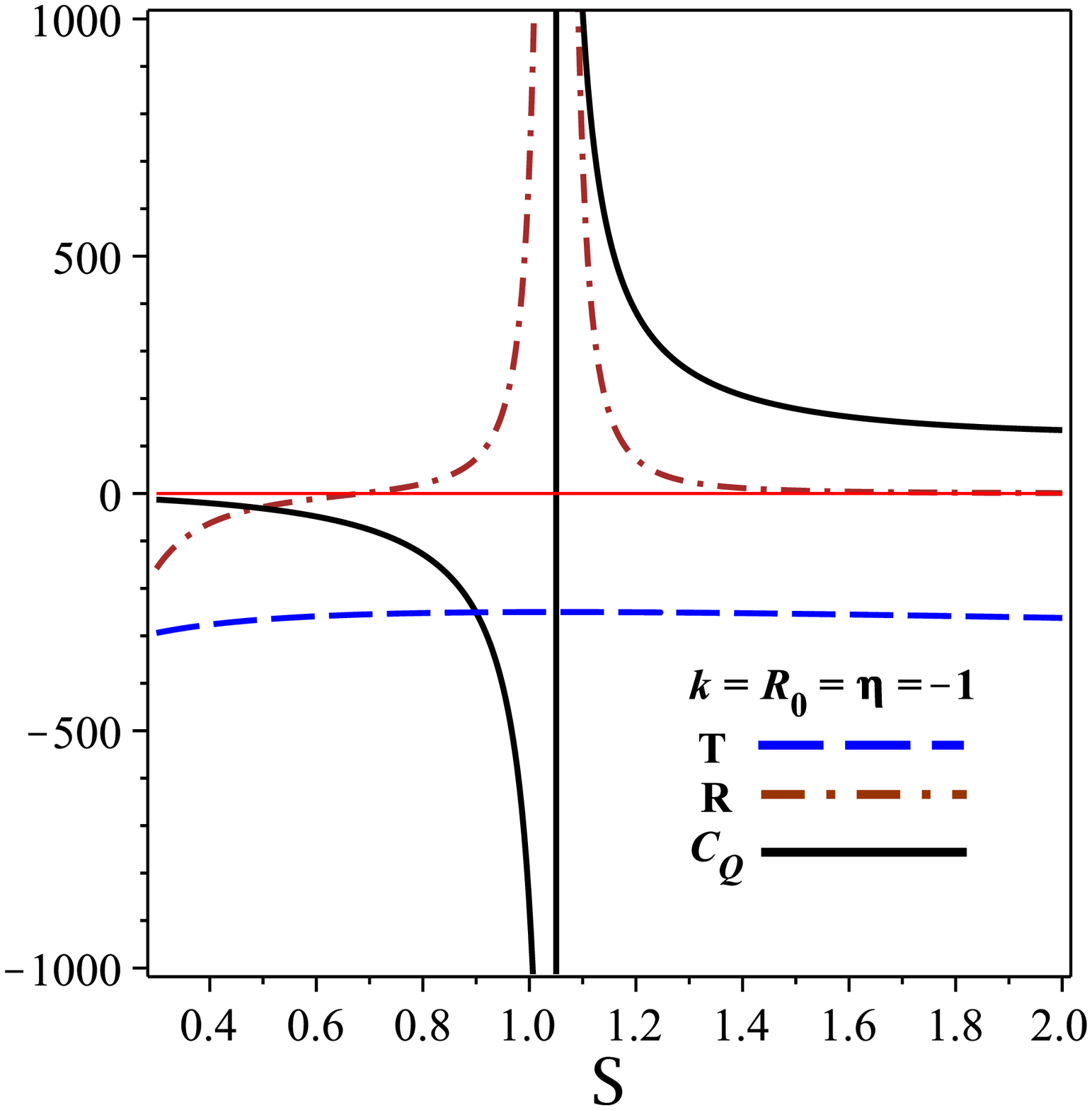}\newline
\caption{The heat capacity ($C_{Q}$) and temperature ($T$) and Ricci scalar (%
$R$) versus $S$ for $Q=0.02$ and $f_{R}=0.1$. We plot them in different
scales to be more clear.}
\label{Fig2b}
\end{figure}

To have a proper geometrothemodynamics approach for studying phase
transitions, the thermodynamic Ricci scalar should diverge at points which
we mentioned before with bound $\left( M=\frac{\partial M}{\partial S}%
=0\right) $ and phase transition $\left( M_{SS}=\frac{\partial ^{2}M}{%
\partial S^{2}}=0\right) $ points (see Eq. (\ref{PhysBound}), for more
details). Regarding Eq. (\ref{RR}), it is evident that the divergence points
and root of the heat capacity coincide with divergences of the HPEM's Ricci
scalar. In other words, the denominator of the Ricci scalar of the HPEM
metric contains the numerator and denominator of the heat capacity (Eq. (\ref%
{Heat})). Indeed the divergence points of the Ricci scalar of the HPEM
metric coincide with both roots and phase transition critical points of the
heat capacity. So, all the physical limitations and the phase transition
critical points are included in the divergences of the Ricci scalar of the
HPEM metric (see Fig. \ref{Fig2b}, for more detail). As a result, the HPEM
metric provides a successful mechanism for investigating the bound and phase
transition points of such black holes.

Looking at the figure. \ref{Fig2b}, we can find another important behavior
of the HPEM metric which is related to the different behavior of the Ricci
scalar before and after its divergence points. The behavior of HPEM's Ricci
scalar for divergence points related to the physical limitation and phase
transition critical points is different. In other words, the sign of HPEM's
Ricci scalar changes before and after divergencies when the heat capacity is
zero. However, the signs of the Ricci scalar are the same when the heat
capacity encounters with divergences. These divergences are called $\Lambda $
divergences. Therefore, considering this approach also enable us to
distinguish the physical limitation and the phase transition critical points
from one another (see Fig. \ref{Fig2b}, for more details).

\subsection{Global Stability}


In the context of the grand-canonical ensemble, the global stability of a
thermodynamic system can be studied by Gibbs's potential. In other words,
the negative of the Gibbs potential determines the global stability of a
thermodynamic system. On the other hand, the negative of the Helmholtz free
energy of a thermodynamic system satisfies the global stability in the
context of the canonical ensemble. Therefore, by using the Gibbs potential
and the Helmholtz free energy, we want to evaluate the global stability of
the topological phantom (A)dS black holes in $F(R)$ gravity.


\subsubsection{Gibbs Potential}


The Gibbs potential is defined in the following form 
\begin{equation}
G=M\left( S,Q\right) -TS-\eta UQ,  \label{Gibbs}
\end{equation}%
by using the relation $\eta U=\left( \frac{\partial M}{\partial Q}\right)
_{S}$ and Eqs. (\ref{MSQ}) and (\ref{TM}), we get the Gibbs potential as 
\begin{equation}
G=\frac{\frac{S}{12}\left( 3\left( 1+f_{R}\right) k-SR_{0}\right) -\pi
^{2}Q^{2}\left( 1+f_{R}\right) \eta }{2\pi \sqrt{S\left( 1+f_{R}\right) }},
\end{equation}%
where the roots of the Gibbs potential are given by 
\begin{equation}
\left\{ 
\begin{array}{c}
S_{G_{1}}=\frac{3\left( 1+f_{R}\right) k}{2R_{0}}-\frac{\sqrt{3\left[
3\left( 1+f_{R}\right) k^{2}-16\pi ^{2}Q^{2}\eta R_{0}\right] \left(
1+f_{R}\right) }}{2R_{0}} \\ 
\\ 
S_{G_{2}}=\frac{3\left( 1+f_{R}\right) k}{2R_{0}}+\frac{\sqrt{3\left[
3\left( 1+f_{R}\right) k^{2}-16\pi ^{2}Q^{2}\eta R_{0}\right] \left(
1+f_{R}\right) }}{2R_{0}}%
\end{array}%
\right. .
\end{equation}

In order to study global stability, we must consider $G<0$. In other words,
the black holes have a global stable when $G<0$. To evaluate the global
stability of the topological phantom (A)dS black holes in $F(R)$ gravity, we
plotted four panels in Fig. \ref{Fig3} (see the panels ($4a$), ($4b$), ($4c$%
), and ($4d$) in Fig. \ref{Fig3}). Our analysis reveals some results which
are:

i) The charged AdS black holes with $k=0$, and $k=-1$ are stable everywhere
(i.e., $S\in \left( 0,+\infty \right) $). For $k=1$, the global stability is
located in the range $S\in \left( 0,S_{G_{1}}\right) \cup \left(
S_{G_{2}},+\infty \right) $, see the panel ($4a$) in Fig. \ref{Fig3}.

ii) The charged dS black holes with small radii and different topological
constants have global stability. In other words, the Gibbs potential is
negative for $S<S_{G}$ (see the panel ($4b$) in Fig. \ref{Fig3}).

iii) The phantom AdS black holes in $F(R)$ gravity with large radii and
different topological constants are stable because the Gibbs potential is
negative for $S>S_{G}$ (see the panel ($4c$) in Fig. \ref{Fig3}).

iv) Whereas the Gibbs potential is negative in the range $S\in \left(
0,S_{G_{1}}\right) \cup \left( S_{G_{2}},+\infty \right) $ for the phantom
dS black holes in $F(R)$ gravity with $k=-1$. In other words, these black
holes are stable in these areas (i.e., $S\in \left( 0,S_{G_{1}}\right) \cup
\left( S_{G_{2}},+\infty \right) $), see the panel ($4d$) in Fig. \ref{Fig3}.


\subsubsection{Helmholtz Free Energy}


Another mechanism for determining the global stability of a thermodynamic
system is related to the Helmholtz free energy. It is notable that in the
usual case of thermodynamics the Helmholtz free energy is given by $F=U-TS$.
However, in the context of the black holes, Helmholtz free energy is defined
in the following form 
\begin{equation}
F(T,Q)=M\left( S,Q\right) -TS,
\end{equation}%
where by considering Eqs. (\ref{MSQ}) and (\ref{TM}), we can obtain the
Helmholtz free energy as 
\begin{equation}
F(T,Q)=\frac{3\pi ^{2}Q^{2}\left( 1+f_{R}\right) \eta -\frac{S}{12}\left(
SR_{0}-3\left( 1+f_{R}\right) k\right) }{2\pi \sqrt{S\left( 1+f_{R}\right) }}%
,
\end{equation}%
and by solving $F(T,Q)=0$, we get the roots of the Helmholtz free energy
that are 
\begin{equation}
\left\{ 
\begin{array}{c}
S_{F_{1}}=\frac{3\left( 1+f_{R}\right) k}{2R_{0}}-\frac{3\sqrt{\left[ \left(
1+f_{R}\right) k^{2}+16\pi ^{2}Q^{2}\eta R_{0}\right] \left( 1+f_{R}\right) }%
}{2R_{0}} \\ 
\\ 
S_{F_{2}}=\frac{3\left( 1+f_{R}\right) k}{2R_{0}}+\frac{3\sqrt{\left[ \left(
1+f_{R}\right) k^{2}+16\pi ^{2}Q^{2}\eta R_{0}\right] \left( 1+f_{R}\right) }%
}{2R_{0}}%
\end{array}%
\right. .
\end{equation}

\begin{figure}[tbph]
\centering
\includegraphics[width=0.4\linewidth]{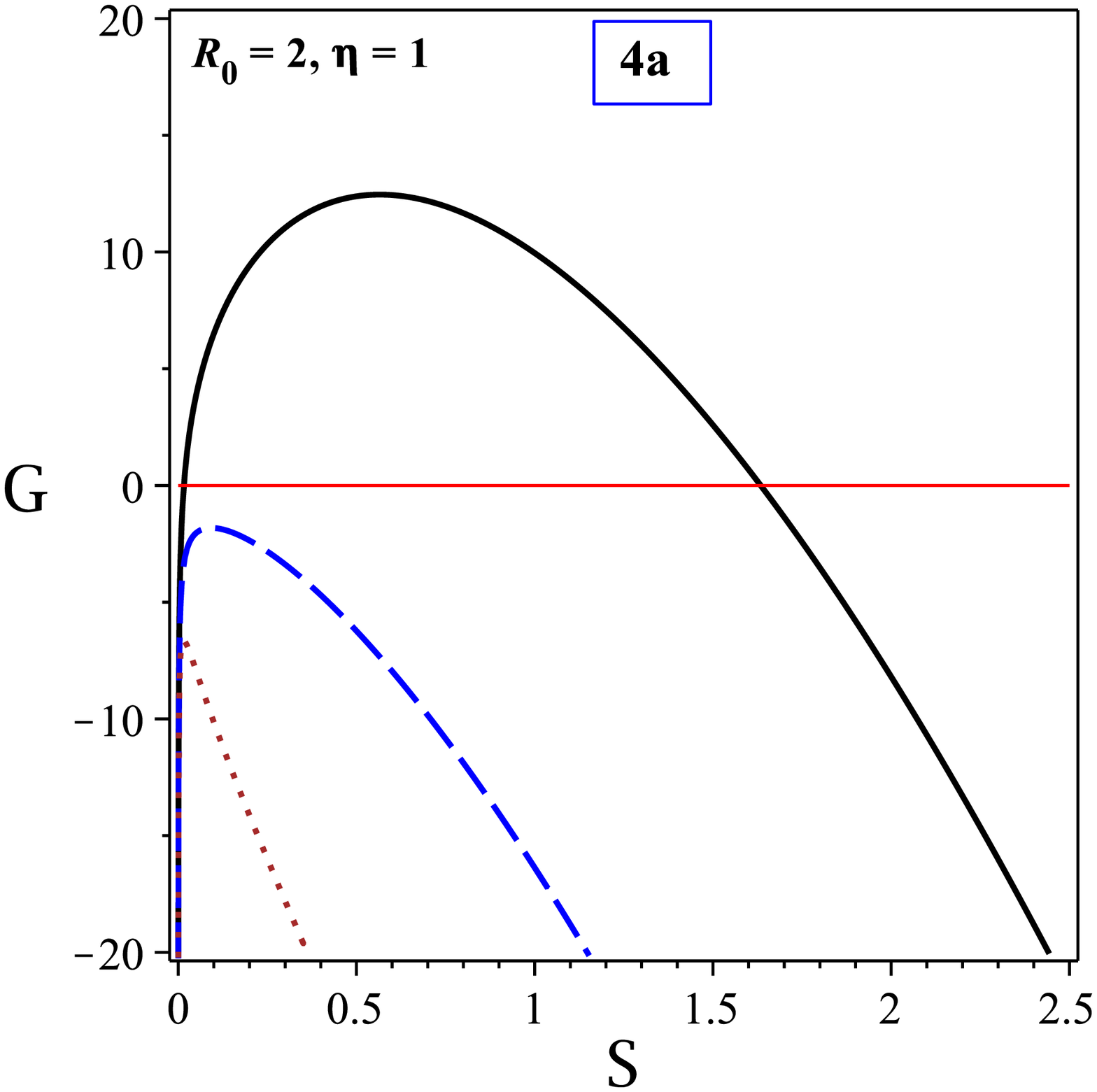} \includegraphics[width=0.4%
\linewidth]{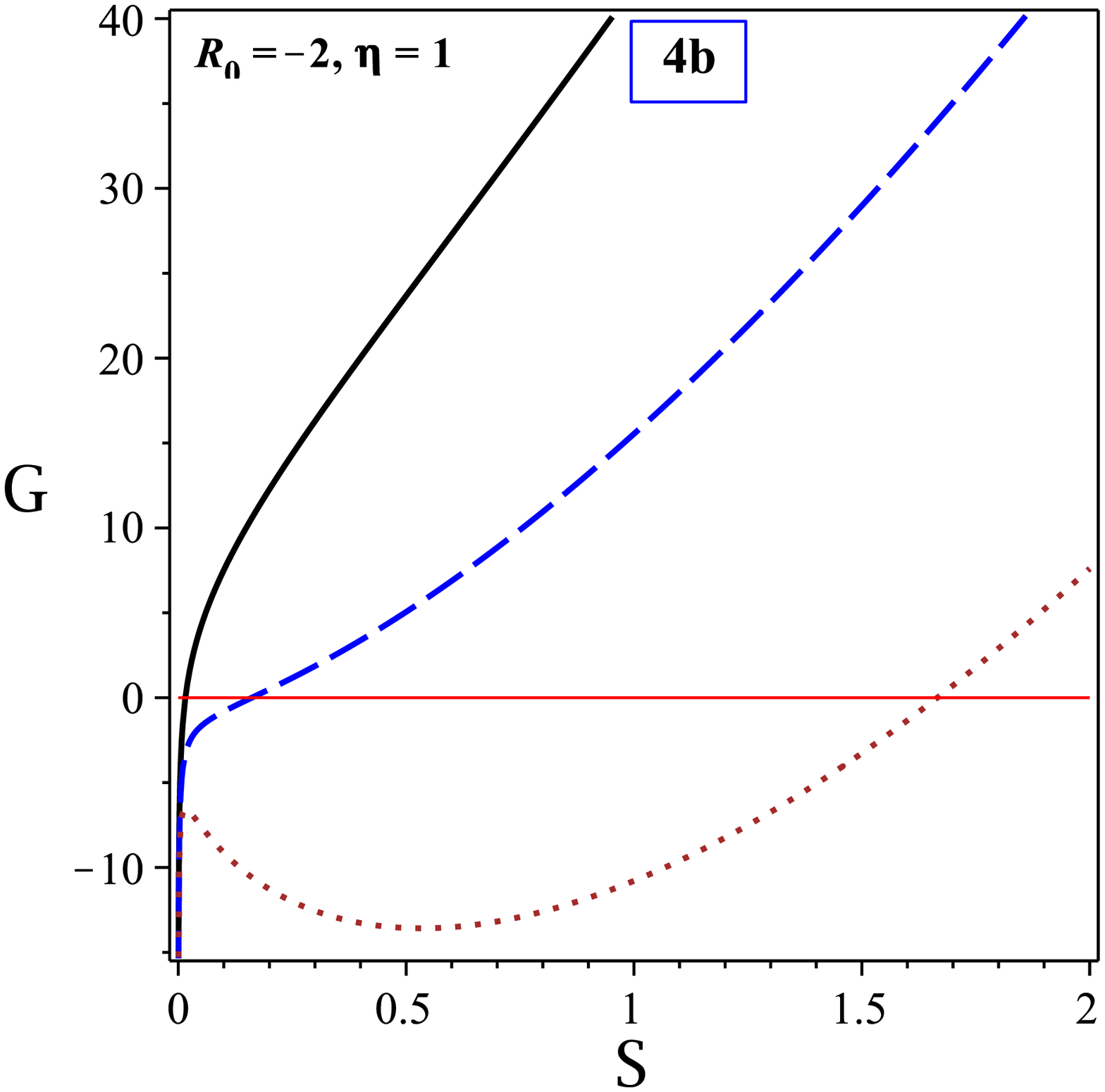} \newline
\includegraphics[width=0.4\linewidth]{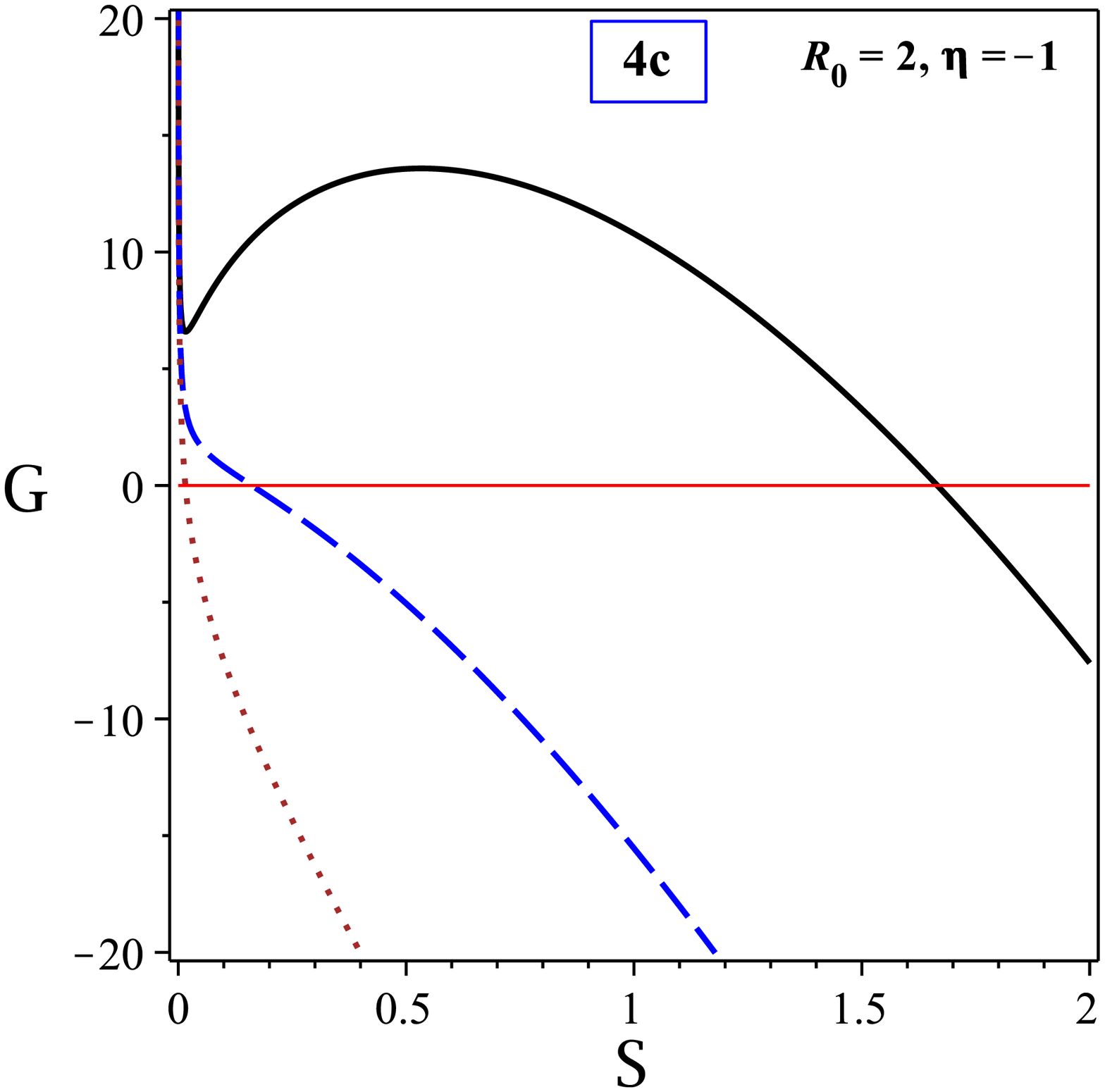} \includegraphics[width=0.4%
\linewidth]{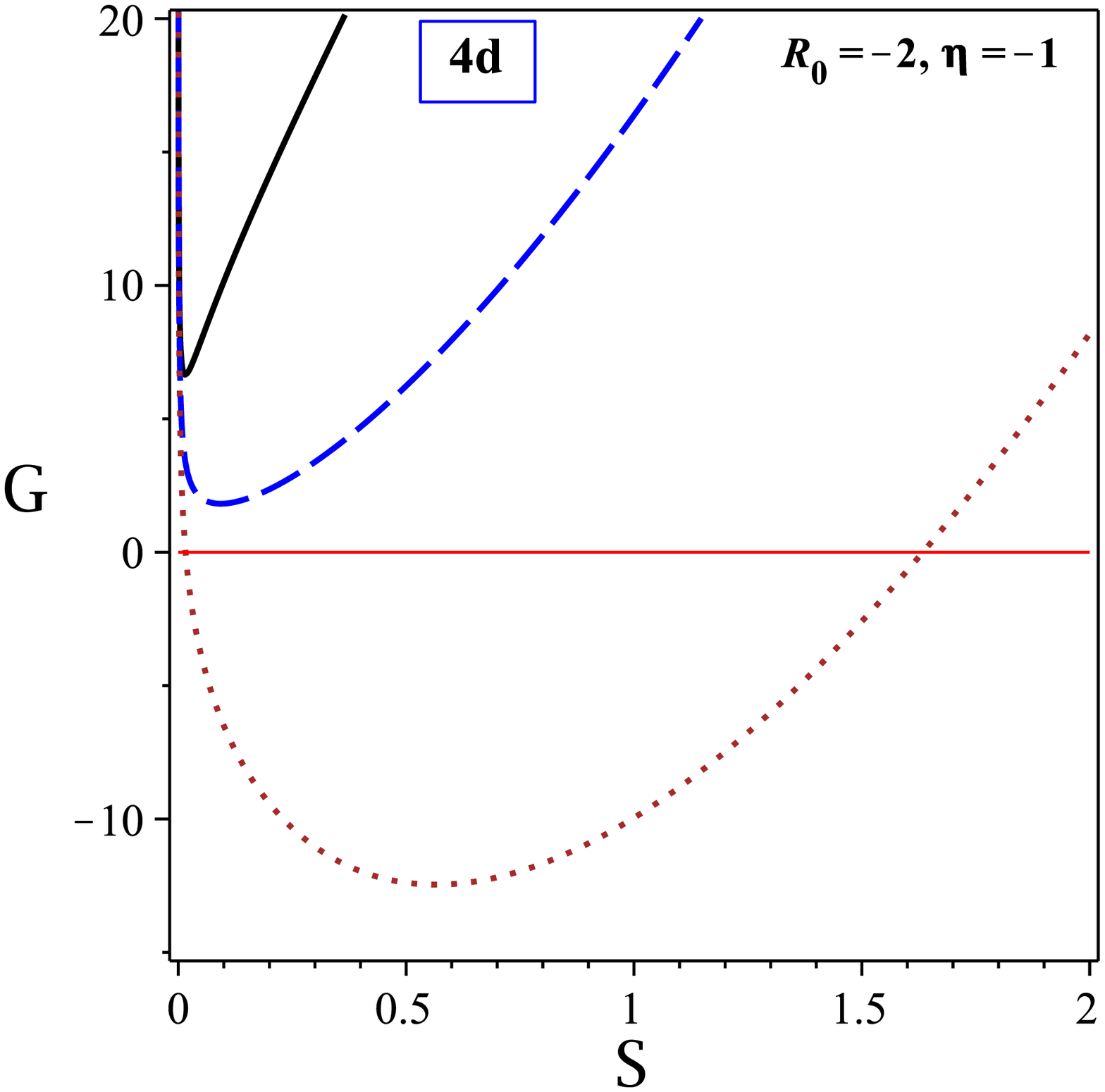} \newline
\includegraphics[width=0.4\linewidth]{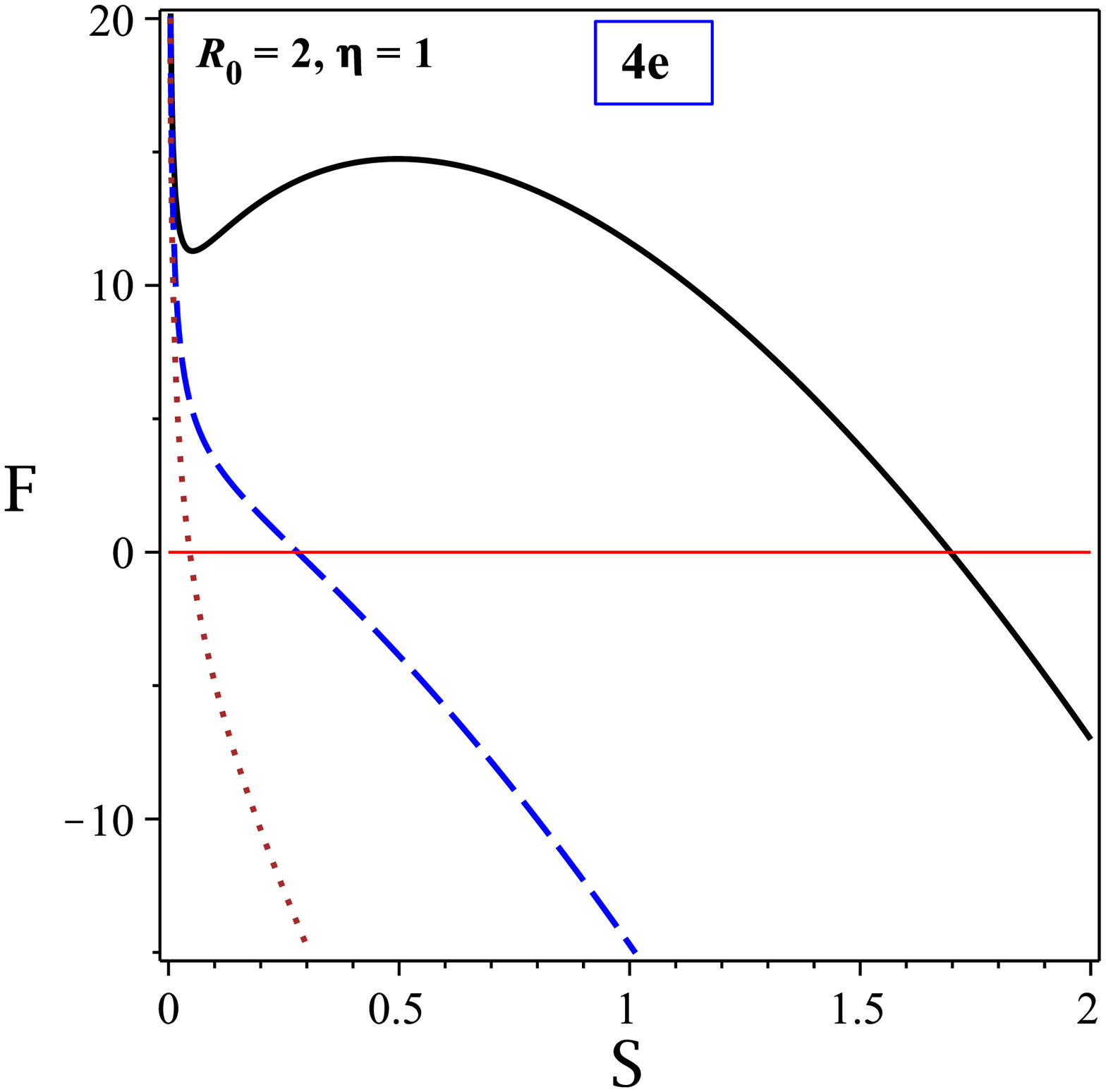} \includegraphics[width=0.4%
\linewidth]{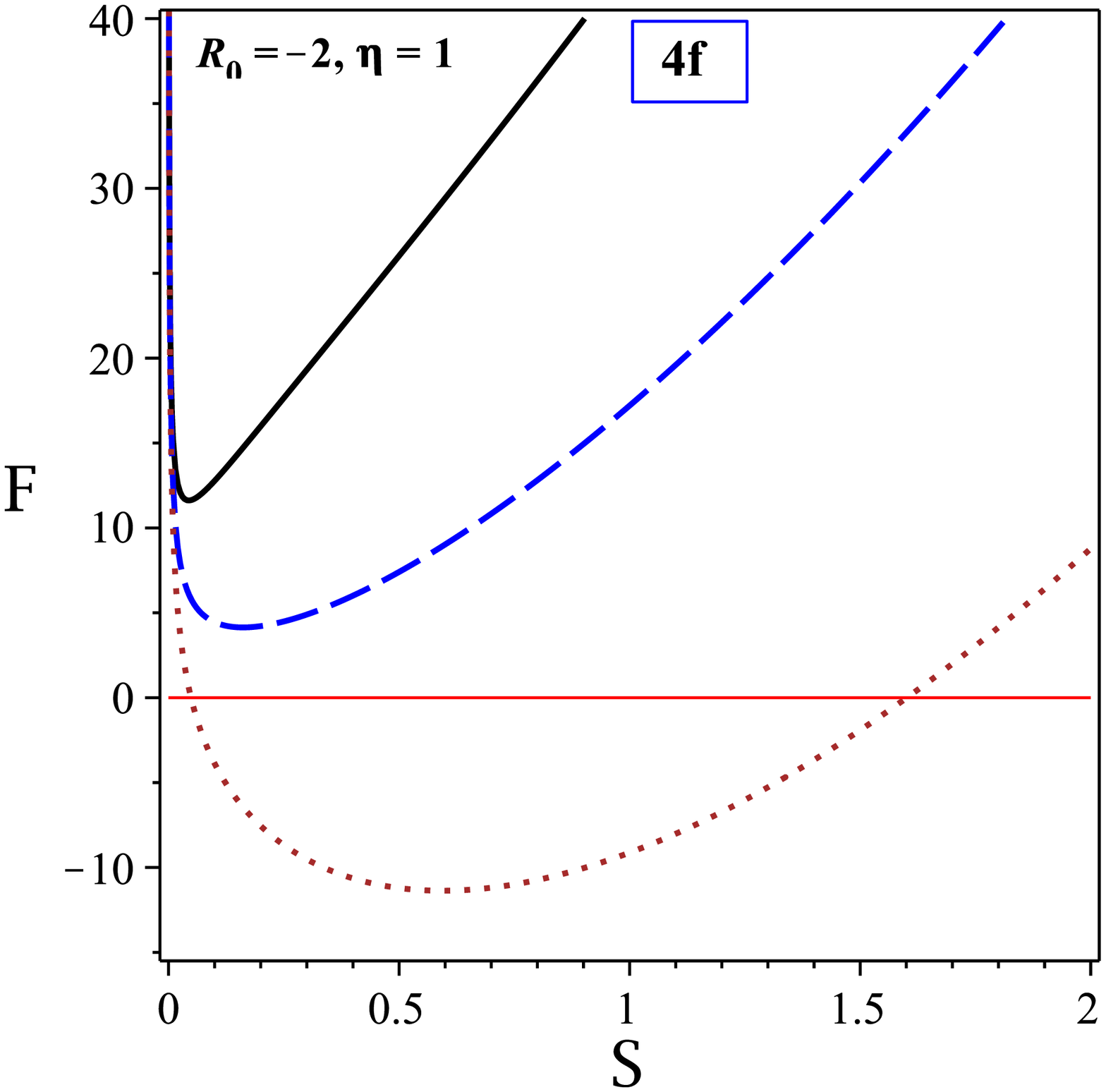}\newline
\includegraphics[width=0.4\linewidth]{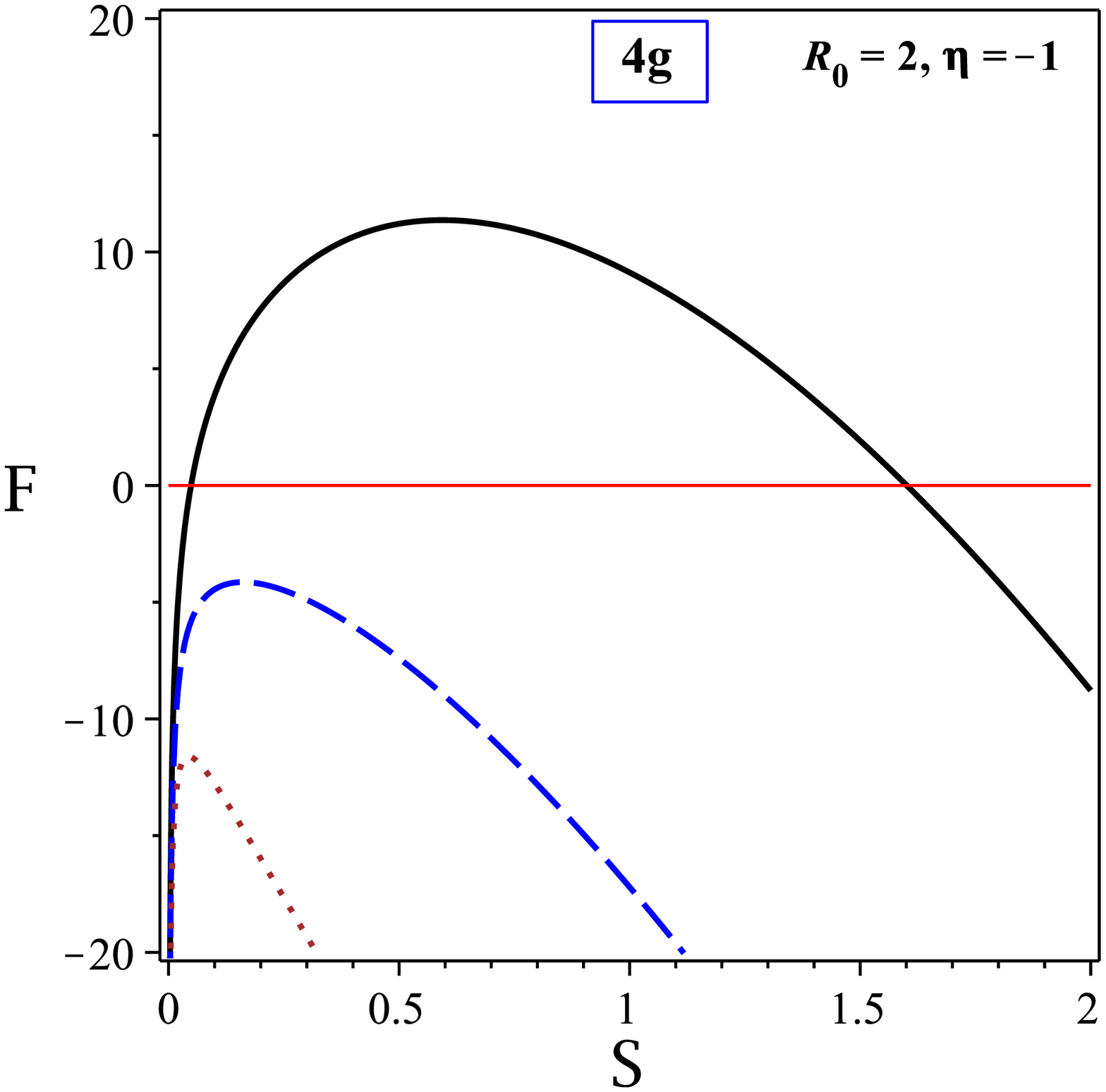} \includegraphics[width=0.4%
\linewidth]{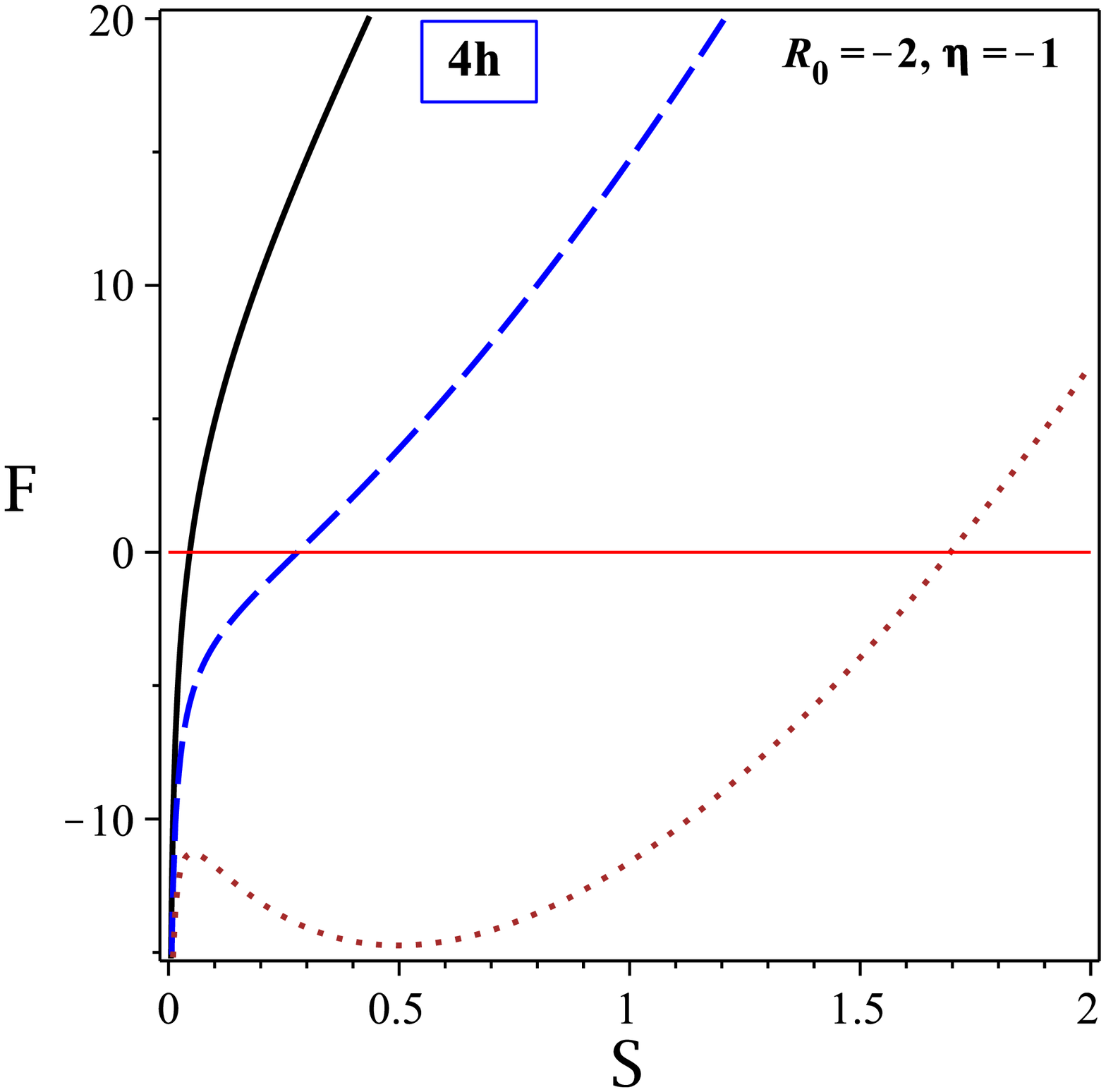}\newline
\caption{$G$ (and $F$) versus $S$ for $Q=0.02$, and $f_{R}=0.1$. Also, $k=1$
(continuous line), $k=0$ (dashed line), and $k=-1$ (dotted line).}
\label{Fig3}
\end{figure}

The global stability areas are given when the Helmholtz free energy is
negative (i.e., $F<0$). To evaluate the global stability of the topological
phantom (A)dS black hole in $F(R)$ gravity, we plot the Helmholtz free
versus $S$ in Fig. \ref{Fig3}. Our results are:

i) The charged AdS black holes with different topological constants are
stable when $S>S_{F_{2}}$. Indeed, the large black holes have global
stability (see the panel ($4e$) in Fig. \ref{Fig3}).

ii) There are only stable areas for the charged dS black holes with $k=-1$
when the entropy is located between two roots (i.e., $S_{F_{1}}<S<S_{F_{2}}$%
), see the panel ($4f$) in Fig. \ref{Fig3}.

iii) The phantom AdS black holes with $k=0$, and $k=-1$\ are stable. Whereas
for $k=1$, the global stability areas are located in the range $S\in \left(
0,S_{G_{1}}\right) \cup \left( S_{G_{2}},+\infty \right) $, see the panel ($%
4g$) in Fig. \ref{Fig3}.

iv) The phantom dS black holes with different topological constants and
small radii are stable (see the panel ($4h$) in Fig. \ref{Fig3}).

\begin{figure}[tbph]
\centering
\includegraphics[width=0.4\linewidth]{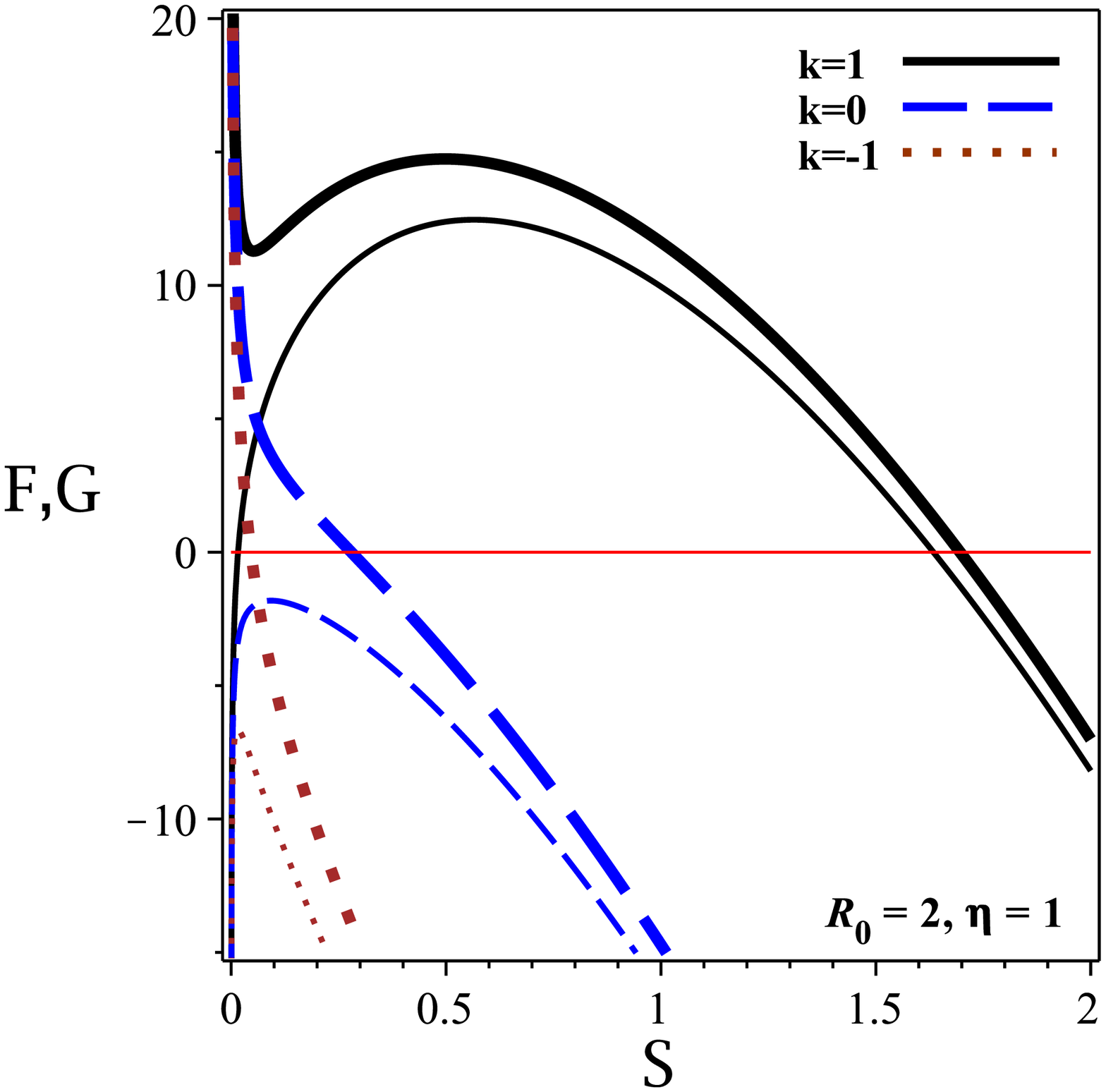} \includegraphics[width=0.4%
\linewidth]{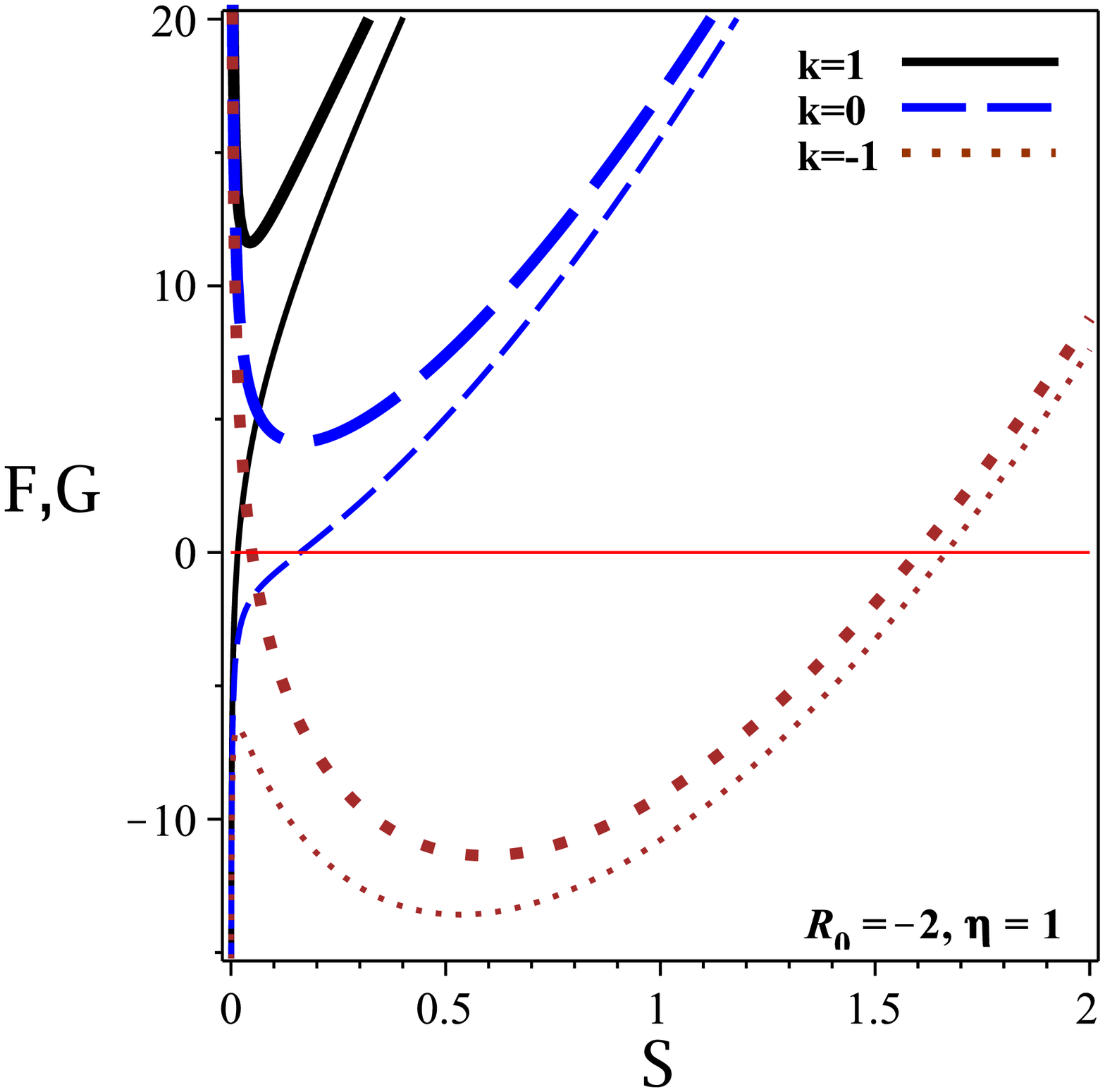} \newline
\includegraphics[width=0.4\linewidth]{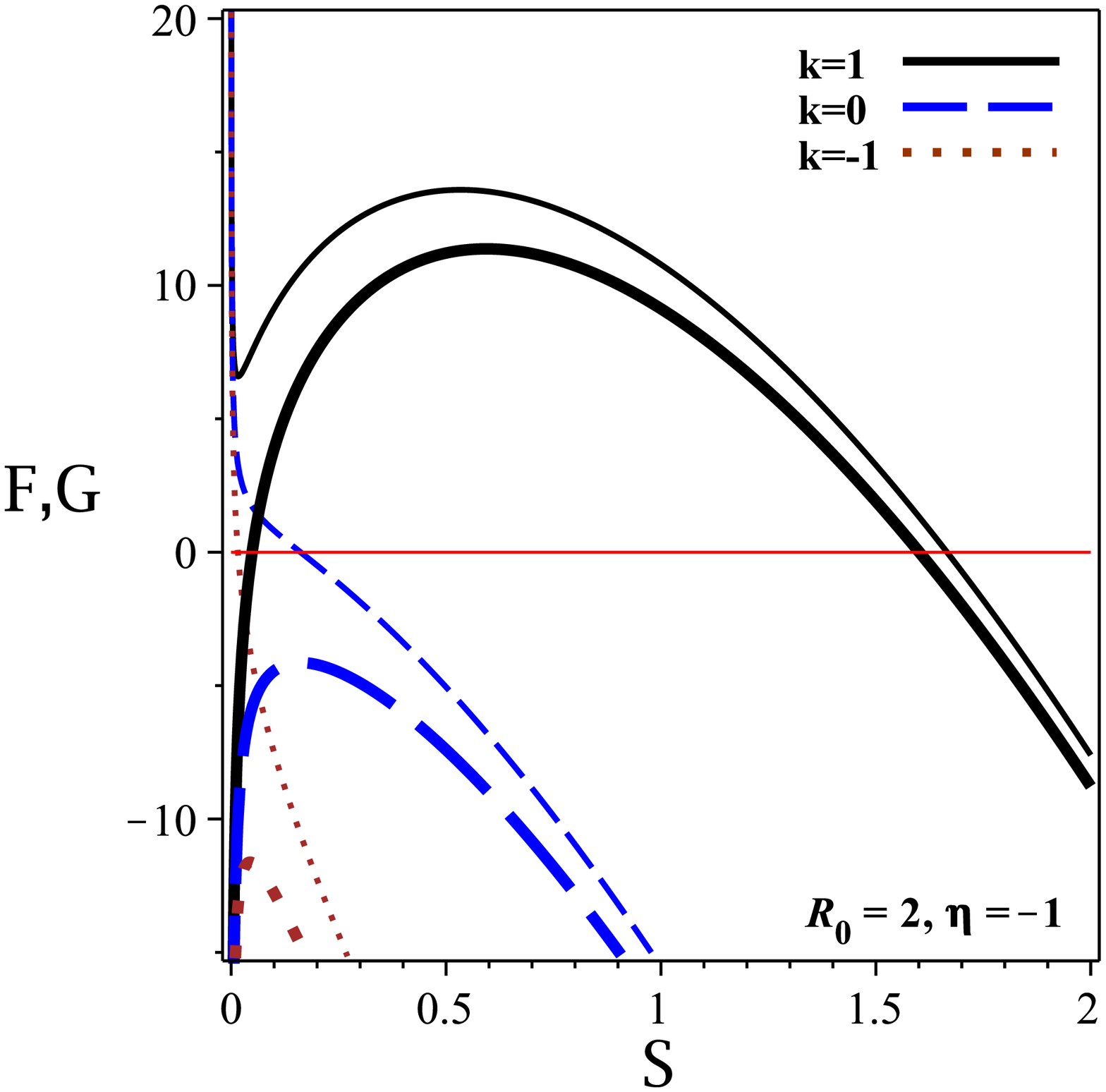} \includegraphics[width=0.4%
\linewidth]{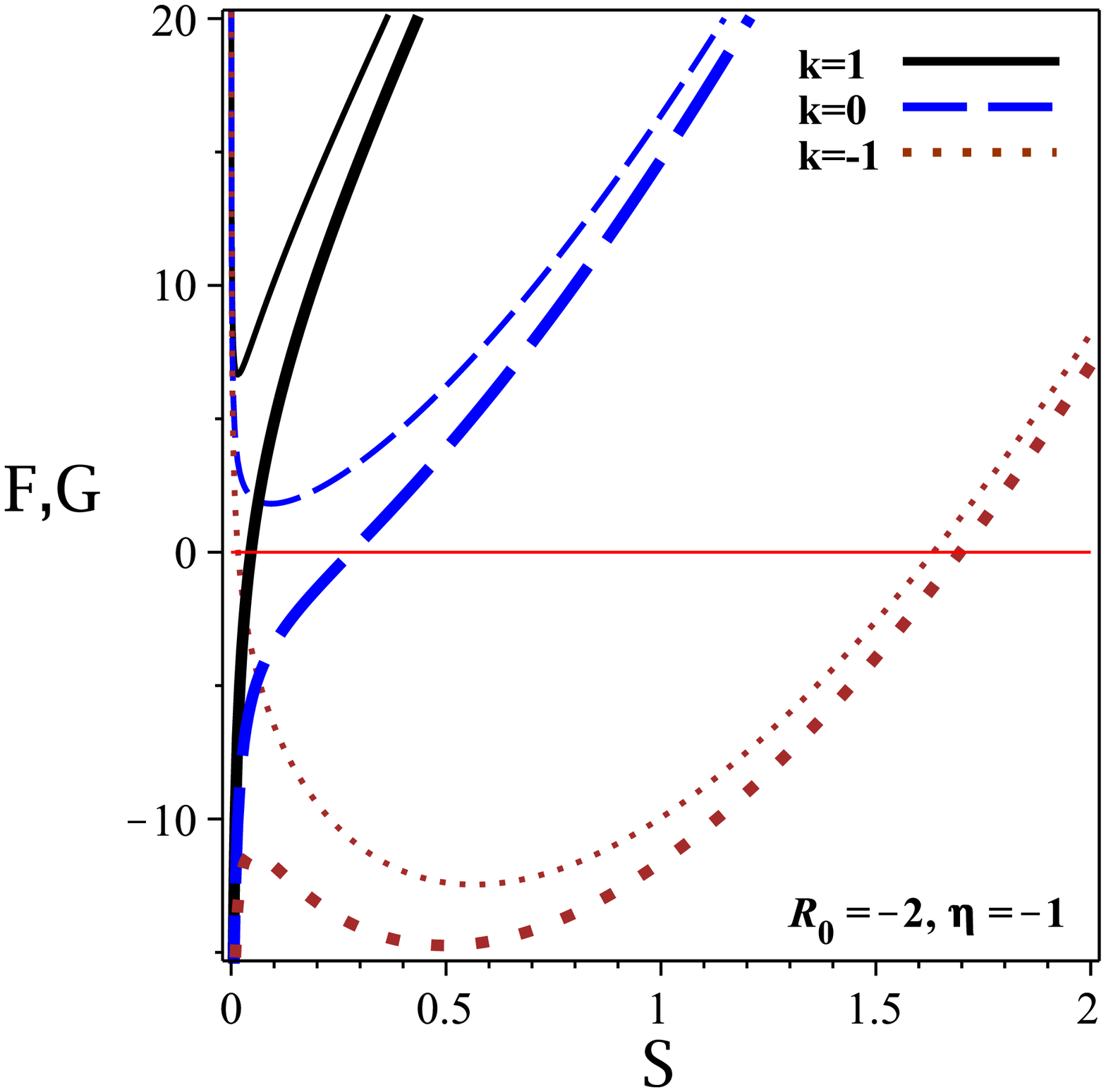} \newline
\caption{The Gibbs potential and the Helmholtz free energy versus $S$ for $%
Q=0.02$, and $f_{R}=0.1$. Thin and bold lines are related to $G$ and $F$,
respectively.}
\label{Fig4}
\end{figure}

In order to have more details of the global stability from two points of
view, we plot the Gibbs potential and the Helmholtz free energy together
versus the entropy in Fig. (\ref{Fig4}).

Comparing these points of view, we find some different behaviors of the
global stability for black holes in $F(R)$ gravity. These results are:

i) Generally, by considering the charged AdS black holes with different
topological constants, the Gibbs potential covers a large global stability
area compared with the Helmholtz free energy. There is the same behavior for
the charged AdS black holes with large radii. Indeed, the large black holes
with $k=0$, $\pm 1$ in two points of view are stable. However, considering
the Gibbs potential, the small charged AdS black holes are stable. Using the
two points of view, the charged AdS black holes with different topological
constants and large radii are always stable (see the left-up panel in Fig. %
\ref{Fig4}, for more details).

ii) Using the Gibbs potential, the charged dS black holes with small radii
and different topological constants are always stable, whereas there is no
stable area for small black holes in the viewpoint of the Helmholtz free
energy. In addition, from point of view of Helmholtz free energy, the
charged dS black holes with medium radii can be stable when $k=-1$. As a
common result from both viewpoints, the charged dS black holes with large
radii and $k=0$, $\pm 1$ do not have the global stability area (see the
right-up panel in Fig. \ref{Fig4}, for more details).

iii) In total, the Helmholtz free energy covers a large global stability
area compared with the Gibbs potential for the phantom AdS black holes with
different topological constants. Our findings indicate that the phantom AdS
black holes with large radii are always stable in the two points of view. On
the other hand, small phantom AdS black holes are only stable in the
viewpoint of the Helmholtz free energy (see the left down panel in Fig. \ref%
{Fig4}, for more details).

iv) The phantom dS black holes with small radii can be always stable in the
viewpoint of Helmholtz free energy. However, from point of view of Gibbs's
potential, the phantom dS black holes with medium radii can be only stable
when $k=-1$ (see the right down panel in Fig. \ref{Fig4}, for more details).


\section{Conclusions}

\label{sec4} 

In section \ref{sec1}, we obtain an exact solution of the case of the theory 
$F(R)$ with constant curvature, for a topological metric in four dimensions,
coupling a spin$-1$ phantom field. We show this solution has one or two
horizons, depending on the value of the topological constant $k$, the mass,
the charge, the coupling constant $\eta $, and the scalar $R_{0}$.

In section \ref{sec2}, we define the temperature, electric field, and
entropy of the solution, as well as establish the first law of
thermodynamics, showing that the term related to thermodynamic work can be
positive or negative, depending on whether the field is phantom or not. This
property was already known in phantom solutions \cite{manuel1}.

In section \ref{sec3}, we first study local thermodynamic stability, through
the zeros and divergent points of the heat capacity, as well as the
geometrothemodynamics, agreeing between the two approaches. Then we study
global thermodynamic stability by analyzing the sign of the Gibbs potential
and Helmholtz free energy. In general, the two approaches agree on slightly
different ranges of entropy.

We should also analyze, in a future work, the geodesics of this solution, as
well as the shadow and stability, checking the absorption and scattering of
scalar fields.

We can find astrophysical evidence of the signature of the phantom
modification of the electromagnetic contribution in the following situations:

a) the shadow of this black hole must present characteristics of
differentiation between the usual Reissner-Nordstr\"{o}m one, thus being
able to serve as experimental evidence.

b) the new phantom signature must also appear in the gravitational
wave ringdown of this solution.

c) observational evidence can also be obtained through gravitational
lensing phenomena.

\begin{acknowledgements}
B. Eslam Panah thanks University of Mazandaran. M. E. R. thanks Conselho Nacional de Desenvolvimento Cientifico e Tecnologico - CNPq, Brazil, for partial financial support.
\end{acknowledgements}

\end{document}